\documentclass[submission, Phys]{SciPost}
\hypersetup{
    colorlinks,
    linkcolor={red!50!black},
    citecolor={blue!50!black},
    urlcolor={blue!80!black}
}

\usepackage[bitstream-charter]{mathdesign}
\usepackage{tikz}
\usepackage{scalerel}
\usepackage{slashed}
\usepackage[b]{esvect}
\usepackage[compat=1.1.0]{tikz-feynman}
\usepackage{caption}
\usepackage{subcaption}

\newcommand{\sslash}{\mathbin{/\mkern-6mu/}}

\DeclareMathOperator*{\argmax}{arg\,max}

\usetikzlibrary{positioning}
\usetikzlibrary{shapes.geometric}
\tikzstyle{node} = [circle,fill=white, draw=black, minimum size=0.7cm]
\tikzstyle{blob} = [circle, minimum size=0.9cm,inner sep=0pt, thin, draw=black]
\tikzstyle{oval} = [ellipse, minimum height=1.5cm, minimum width=0.1cm, inner sep=0pt, thin, draw=black]
\tikzset{arrow/.style={-stealth, semithick, draw=black}}
\tikzset{dashedarrow/.style={-stealth, semithick, draw=black}}
\usetikzlibrary{decorations.pathreplacing}
\usetikzlibrary{shapes.multipart}

\DeclareSymbolFont{usualmathcal}{OMS}{cmsy}{m}{n}
\DeclareSymbolFontAlphabet{\mathcal}{usualmathcal}

\begin{document}

\begin{center}{\Large \textbf{
Event Generation and Density Estimation with Surjective Normalizing Flows
}}\end{center}

% TODO: write the author list here. Use first name (+ other initials) + surname format.
% Separate subsequent authors by a comma, omit comma and use "and" for the last author.
% Mark the corresponding author with a superscript star.
\begin{center}
Rob Verheyen\textsuperscript{1$\star$}
\end{center}

\begin{center}
{\bf 1} Department of Physics and Astronomy
University College London, Gower St., Bloomsbury, London WC1E 6BT, UK
\\
${}^\star$ {\small \sf r.verheyen@ucl.ac.uk}
\end{center}

\begin{center}
\today
\end{center}

% For convenience during refereeing (optional),
% you can turn on line numbers by uncommenting the next line:
%\linenumbers

\section*{Abstract}
{\bf Normalizing flows are a class of generative models that enable exact
likelihood evaluation. While these models have already found various
applications in particle physics, normalizing flows are not flexible enough to
model many of the peripheral features of collision events. Using the framework
of \cite{nielsen2020survae}, we introduce several surjective and stochastic
transform layers to a baseline normalizing flow to improve modelling of
permutation symmetry, varying dimensionality and discrete features, which are
all commonly encountered in particle physics events. We assess their efficacy in
the context of the generation of a matrix element-level process, and in the
context of anomaly detection in detector-level LHC events. }

\vspace{10pt}
\noindent\rule{\textwidth}{1pt}
\tableofcontents\thispagestyle{fancy}
\noindent\rule{\textwidth}{1pt}
\vspace{10pt}

\section{Introduction}
First-principle Monte Carlo event generators are a fundamental component of most
LHC physics analyses. As the LHC enters its third run, and with the high
luminosity upgrade in the near future, the amount of available experimental data
is set to increase rapidly. To match the resulting statistical precision, the
event generators must follow suit. The nature of perturbative calculations in quantum
field theory is that such an increase in precision of the simulations comes
hand-in-hand with an increase in complexity, and thus with more costly
simulations. This means that advances in event generator technology are required to
maintain interpretability of future LHC data
\cite{Buckley:2019wov,HSFPhysicsEventGeneratorWG:2020gxw}.

One promising avenue to tackle these computational challenges can be found in
modern machine learning techniques. In particular, generative models such as
generative adversarial networks (GANs) \cite{NIPS2014_5ca3e9b1}, variational
autoencoders (VAEs) \cite{kingma2013auto} and normalizing flows
\cite{10.5555/3045118.3045281,DBLP:journals/corr/abs-1808-04730} offer alternatives for fast event generation. Of
these models, normalizing flows have the particular advantage of simultaneously
enabling event generation and likelihood evaluation, the latter of which is
useful in other applications. As a result, they have been successfully used for
a variety of tasks including event generation
\cite{Bothmann:2020ywa,Gao:2020zvv,Stienen:2020gns,Butter:2021csz,Butter:2022lkf,Krause:2021ilc,Krause:2021wez,Jawahar:2021vyu,Hollingsworth:2021sii,Choi:2020bnf,Gao:2020vdv,Brehmer:2020vwc,Bellagente:2021yyh},
anomaly detection \cite{Caron:2021wmq,Hallin:2021wme,Nachman:2020lpy}, unfolding
\cite{Bellagente:2020piv}, the calculation of loop integrals
\cite{Winterhalder:2021ngy}, and likelihood-free inference
\cite{Bieringer:2020tnw,Vandegar:2020yvw,Shirobokov:2020tjt}.

Normalizing flows make use of a set of differentiable bijective functions to
transform between a simple, fixed base distribution and a complex, learned
distribution. Much progress has been made in the development of normalizing flow
architectures that are both expressive and efficient
\cite{dinh2014nice,dinh2016density,kingma2017improving,papamakarios2017masked,huang2018neural,kingma2018glow,jaini2019sum,jaini2020tails,durkan2019neural}.
However, normalizing flows lack in flexibility due to the bijective nature of
the transforms, meaning that they are essentially restricted to modelling a
continuous feature space of fixed dimension. For the purposes of event
generation and likelihood estimation in particle physics, more flexibility is
often required, for instance to model discrete features or varying
dimensionality on an event-by-event basis.

On the other hand, GANs and VAEs do not have these limitations, but they do not
offer exact likelihood evaluation. GANs are trained adversarially and thus do
not offer access to the likelihood at all, but VAEs are able to provide a lower
bound estimate of the likelihood. In \cite{nielsen2020survae} a method was
outlined that combines the favorable properties of normalizing flows with the
flexibility of VAEs. In this work, we explore the use of this framework, as well
as other solutions, to improve the flexibility of normalizing flows in the
context of particle physics event generation and density estimation.

In section \ref{sec:latent-variable-models}, we summarize normalizing flow and
VAE architectures, as well as their combination as detailed in
\cite{nielsen2020survae}. Section \ref{sec:gluino} describes a test case and a
baseline normalizing flow, which are then used to explore the incorporation of
permutation invariance (section \ref{sec:perm-invariance}), varying
dimensionality (section \ref{sec:varying-dim}) and discrete features (section
\ref{sec:discrete}). In section \ref{sec:DM} these techniques are then applied
to a density estimation problem in the context of the Dark Machines Anomaly
Score Challenge \cite{Aarrestad:2021oeb}. We conclude in section
\ref{sec:conclusions}.

\section{Surjective Normalizing Flows} \label{sec:latent-variable-models}

We are interested in setting up a generative model that is able to generate new
events, but also evaluate the likelihood of existing events. \emph{Latent
variable models} are one such class of models. They are typically composed of a
set of relatively simple components, but turn out to be expressive enough to
learn the complicated probability distributions that are commonly encountered in
particle physics. Given a set of physical events $x \in \mathcal{X}$ of
dimension $d_x$, we define an auxiliary set of latent variables $z \in
\mathcal{Z}$ of dimension $d_z$ with an associated joint probability
distribution $p(x,z)$, which is specified by the model and thus depends on a set
of trainable parameters. The marginal probability density 
\begin{equation} \label{eq:latent-variable-model}
    p(x) = \int_{\mathcal{Z}} dz \, p(x,z) = \int_{\mathcal{Z}} dz \, p(z) \, p(x|z)
\end{equation}
is then the distribution of interest. The second equality in
eq.~\eqref{eq:latent-variable-model} arises through the general product rule,
and implies a generative process, which is given by 
\begin{align} \label{eq:latent-sampling}
    z &\sim p(z) \nonumber \\
    x &\sim p(x|z).
\end{align}
This generative process is efficient as long as $p(z)$ and $p(x|z)$ are simple
enough to be sampled from, while the conditioning on the latent variables $z$
leads to increased expressivity. However, the trade-off is that the evaluation
of the marginal likelihood, eq.~\eqref{eq:latent-variable-model}, is generally
not tractable. Aside from the fact that likelihood evaluation is an objective in
its own right, the training of probabilistic models is generally accomplished by
maximum likelihood estimation, or equivalently, minimization of the
Kullback-Leibler divergence $\mathbb{D}_{\text{KL}}$ with an empirical
distribution $p_{\text{data}}(x)$,
i.e.
\begin{align} \label{eq:mle}
    \mathcal{L}_{\text{MLE}} &= \mathbb{E}_{p_{\text{data}}(x)}\big[-\log p(x) \big] \nonumber \\
    &= \mathbb{D}_{\text{KL}} \big[p_{\text{data}}(x) | p(x) \big] - \underbrace{\int_{\mathcal{X}} dx \, p_{\text{data}}(x) \log p_{\text{data}}(x)}_{\text{constant}},
\end{align}
where $\mathbb{E}_{p_{\text{data}}(x)}$ indicates an expectation value over
$p_{\text{data}}(x)$. Eq.~\eqref{eq:mle} also requires likelihood evaluation,
and as such the intractability of eq~\eqref{eq:latent-sampling} is a significant
issue, for which several solutions exist. 

\subsection{Normalizing Flows}
\begin{figure}
    \centering
    \begin{tikzpicture}
        % Sort figure

        % Nodes
        \node at (-1.25,0.0)(pz0) {$p(z_0) \sim$};
        \node[blob] at (0.0,0.0)(z0) {$z_0$};
        \node[blob] at (2.0,0.0)(z1) {$z_1$};
        \node[blob] at (4.0,0.0)(z2) {$z_2$};
        \node[blob] at (6.0,0.0)(x) {$x$};
            
        % Arrows between zs
        \draw[arrow] (z0) to [out=45,in=135] (z1);
        \draw[arrow] (z1) to [out=225,in=315] (z0);
        \draw[arrow] (z1) to [out=45,in=135] (z2);
        \draw[arrow] (z2) to [out=225,in=315] (z1);
        \draw[arrow] (z2) to [out=45,in=135] (x);
        \draw[arrow] (x) to [out=225,in=315] (z2);
        
        \fill[white](4.9,0.5) rectangle(5.1,0.7);
        \fill[white](4.9,-0.5) rectangle(5.1,-0.7);
        \node at (5.0,0.6) {\footnotesize $\sslash$};
        \node at (5.0,-0.6) {\footnotesize $\sslash$};

        % Function transforms
        \node at (1.0, 0.0) {$f_1$};
        \node at (3.0, 0.0) {$f_2$};
        \node at (5.,0) {...};
    \end{tikzpicture}
    \caption{ Visualization of a normalizing flow architecture. The forward
        direction starts with a sample $z_0$ distributed according to the base
        distribution $p(z_0)$ after which $n$ flow bijections $f_i$ are applied
        to arrive at $z_n = x$. In the inverse direction, starting from $x$, the
        flow transforms are applied in reverse order to arrive at $z_0$, for
        which $p(z_0)$ can be evaluated. } 
    \label{fig:flow-illustration}
\end{figure}
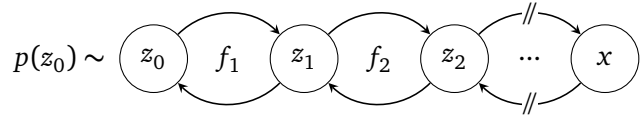
One option to resolve the intractability of eq.~\eqref{eq:latent-variable-model}
is to remove the stochastic component from the conditional probability
distribution, $p(x|z) = \delta(x - f(z))$, leading to (in log-space)
\begin{align} \label{eq:flow-step-llh}
    \log p(x) &= \log\bigg[ \int_{\mathcal{Z}} dz \, p(z) \, \delta(x - f(z)) \bigg] = \log p(z) + \log |J(x)|,
\end{align}
where $|J(x)|$ is the Jacobian determinant associated with the transform $f(z)$.
The Dirac delta function requires $d_x = d_z$, and the evaluation of $p(z) =
p(f^{-1}(x))$ is only possible if $f(z)$ is a bijective function.
Eq.~\eqref{eq:flow-step-llh} is the fundamental step in \emph{normalizing flow}
architectures. Normalizing flow transforms are composable, meaning that to
further improve expressivity, multiple may be stacked, $z_0 \to z_1 \to ... \to
z_n = x$, leading to 
\begin{equation} \label{eq:flow-llh}
    \log p(x) = \log p(z_0) + \sum_{i=1} \log |J_i(z_i)|.
\end{equation}
The base distribution $p(z)$ can then be taken to be simple, such as a
multivariate uniform or a standard normal with diagonal covariance. The
generative process then involves drawing a sample from $p(z)$, which is then
passed through the layers of flow transforms in the \emph{forward} direction
until reaching $x$. On the other hand, likelihood evaluation starts from $x$
which is passed in the \emph{inverse} direction until $z$ is reached while
aggregating the Jacobian determinant of every transform along the way. The prior
can then be evaluated and eq.~\eqref{eq:flow-llh} can be computed. Figure
\ref{fig:flow-illustration} shows an illustration of the normalizing flow
architecture.

Much of the research on normalizing flows has
focussed on improving the expressiveness and efficiency of the bijective
transform, see e.g. \cite{KobyzevPAMI2020,papamakarios2021normalizing} for
reviews. We describe the specific architecture used in this work in section
\ref{sec:gluino}.

\subsection{Variational Inference}
Instead of solving the intractability of eq.~\eqref{eq:latent-variable-model} by
constraining $p(x|z)$ to a delta function, another option is \emph{variational
inference}\footnote{The objective of variational inference is often stated as
the computation of the posterior $p(z|x) = p(z,x)/p(x)$, for which the
marginalized distribution $p(x)$ is also required.}. In that case, one
introduces a variational approximation $q(z|x)$ to the true posterior $p(z|x)$.
The log-likelihood may then be rewritten as 
\begin{align} \label{eq:elbo-1}
    \log p(x) &= \int_{\mathcal{Z}} dz \, q(z|x) \, \log \frac{p(x|z) p(z)}{p(z|x)} \nonumber \\
    &= \int_{\mathcal{Z}} dz \, q(z|x) \, \bigg[ \log p(x|z) - \log \frac{q(z|x)}{p(z)} + \log \frac{q(z|x)}{p(z|x)} \bigg]\nonumber \\
    &= \mathbb{E}_{q(z|x)} \big[\log p(x|z)\big] - \mathbb{D}_{\text{KL}}\big[q(z|x), p(z)\big] + \mathbb{D}_{\text{KL}}\big[q(z|x), p(z|x)\big],
\end{align}
where in the first line we have added a factor $\int_{\mathcal{Z}} dz \, q(z|x)
= 1$ since $p(x)$ does not depend on $z$, and then used Bayes rule to rewrite
$p(x)$. The intractability of $\log p(x)$ is now isolated in the third term of
eq.~\eqref{eq:elbo-1}, which is strictly positive. The combination of the other
two terms is commonly referred to as the evidence lower bound (ELBO). Due to the
positivity of the third term, the ELBO can be optimized in place of the full
likelihood. Eq.~\eqref{eq:elbo-1} serves as the foundation of the VAE, in which
$p(x|z)$ and $q(z|x)$ are parameterized by deep neural networks and during
training the ELBO is evaluated with a single Monte Carlo sample $z \sim q(z|x)$.
Contrary to the normalizing flow approach, variational autoencoders do not
require any restrictions on the form of $p(x|z)$. However, the gap between the
likelihood and the ELBO vanishes only in the limit where $q(z|x) = p(z|x)$,
which in practice is difficult to accomplish.

\subsection{Surjective and Stochastic Transforms} \label{sec:surflows}
In \cite{nielsen2020survae} it was pointed out that the normalizing flow and VAE
paradigms can the unified by rewriting eq.~\eqref{eq:elbo-1} as 
\begin{align} \label{eq:elbo-2}
    \log p(x) &= \mathbb{E}_{q(z|x)} \bigg[ \log p(z) + \underbrace{\log \frac{p(x|z)}{q(z|x)}}_{\mathcal{V}(x,z)} + \underbrace{\log \frac{q(z|x)}{p(z|x)}}_{\mathcal{E}(x,z)} \bigg],
\end{align}
where $\mathcal{V}(x,z)$ is the \emph{likelihood contribution} and
$\mathcal{E}(x,z)$ is the \emph{bound looseness}. For a normalizing flow
transform, no variational approximation of the posterior is required, i.e.
$p(x|z) = \delta(x - f(z))$ and $q(z|x) = \delta(z - f^{-1}(x))$. The result is
that $\mathcal{V}(x,z) = \log |J(x)|$ and $\mathcal{E}(x,z) = 0$, recovering
eq.~\eqref{eq:flow-step-llh}. However, for stochastic transforms like that of
the VAE, $\mathcal{V}(x,z)$ may be evaluated with a single Monte Carlo sample,
while $\mathcal{E}(x,z)$ remains intractable, again serving as a (strictly
positive) error on the full likelihood. 

Furthermore, it is possible to define \emph{surjective} transforms, which are
deterministic in one direction and stochastic in the other. In case of a
surjection in the inverse direction $x \to z$, i.e. $q(z|x) = \delta(z - g(x))$
but $p(x|z)$ remains stochastic, the bound looseness vanishes if $p(x|z)$ only
has support over the set $B(z) = \{x | z = g(x)\}$.\footnote{In this case, the
posterior $p(z|x) = \delta(z - g(x))$ because any value of $x$ can only have
originated from $z = g(x)$. As a result, $q(z|x) = p(z|x)$ and $\mathcal{E}(x,z)
= 0$.} In case of a surjection in the forward direction $z \to x$ however,
$p(x|z) = \delta(x - h(z))$ and $q(z|x)$ stochastic, the bound looseness is
nonzero.

Section \ref{sec:gluino} will explore several of these transforms, as they
will turn out to be useful in the modelling of several features commonly
encountered in particle collision events. Note that eq.~\eqref{eq:elbo-2}
naturally supports the composable nature of a normalizing flow akin to
eq.~\eqref{eq:flow-llh}, such that bijective, surjective and stochastic
transforms may be combined.

\section{Application in Particle Physics Events} \label{sec:gluino} In this
section we explore the use of surjective transforms as part of a normalizing
flow to improve the handling of several distinctive features of particle physics
events: permutation invariance, varying dimensionalities and discrete features.
We first describe a relatively low-dimensional benchmark process which displays
all of these features, and determine a baseline normalizing flow architecture
that is used throughout, before continuing with a description of techniques and
an assessment of their efficacy.

\subsection{A Benchmark Process} \label{sec:benchmark}
\begin{figure}
    \centering
    \begin{tikzpicture}
        \begin{feynman}
            \vertex (g1) at (0,2);
            \vertex (g2) at (0,-2);
            \vertex (g3) at (1,0);
            \vertex (g4) at (2,0);
            \vertex (g5) at (3,1);
            \vertex (g6) at (3,-1);
            \vertex (g7) at (4,2);
            \vertex (g8) at (4,0);
            \vertex (g9) at (5,1);
            \vertex (g10) at (5,-1);

            \diagram* {
                (g1) -- [gluon] (g3) -- [gluon] (g2), 
                (g5) -- (g4) -- (g6),
                (g5) -- [gluon] (g4) -- [gluon] (g6),
                (g3) -- [gluon] (g4),
                (g5) -- [gluon] (g8),
                (g5) -- [gluon] (g7),
                (g5) -- (g7),
                (g9) -- [gluon] (g8) -- [gluon] (g10),
                (g9) -- (g8) -- (g10)
            };

            \vertex (h1) at (6,2);
            \vertex (h2) at (6,-2);
            \vertex (h3) at (8.5,2);
            \vertex (h4) at (8.5,-2);
            \vertex (h5) at (8.5,0.666666666);
            \vertex (h6) at (8.5,-0.666666666);
            \vertex (h7) at (11,2);
            \vertex (h8) at (11,0.666666666);
            \vertex (h9) at (11,-0.666666666);
            \vertex (h10) at (11,-2);

            \diagram* {
                (h1) -- [gluon] (h3),
                (h2) -- [gluon] (h4),
                (h7) -- [gluon] (h3) -- [gluon] (h5) -- [gluon] (h8),
                (h7) -- (h3) -- (h5) -- (h8),
                (h9) -- [gluon] (h6) -- [gluon] (h4) -- [gluon] (h10),
                (h9) -- (h6) -- (h4) -- (h10),
                (h5) -- [gluon] (h6)
            };
        \end{feynman}
    \end{tikzpicture}
    \caption{Example Feynman diagrams that contribute to the $g \kern 0.05em g \rightarrow \tilde{g} \kern 0.05em \tilde{g} \kern 0.05em \tilde{g} \kern 0.05em \tilde{g}$ matrix element.}
    \label{fig:feyndiag}
\end{figure}
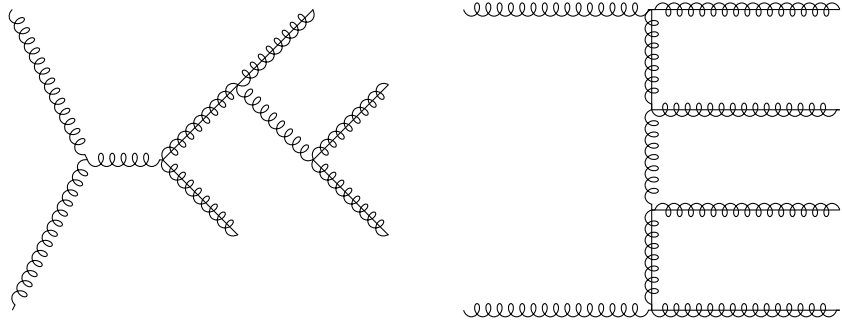

We consider the matrix element-level process 
\begin{equation} \label{eq:process}
    g \kern 0.05em g \rightarrow \tilde{g} \kern 0.05em \tilde{g} \kern 0.05em \tilde{g} \kern 0.05em \tilde{g},
\end{equation}
at $3$ TeV, using the default parameters of the \texttt{MSSM\textunderscore
SLHA} model of \texttt{Madgraph5\textunderscore aMC@NLO} \cite{Alwall:2014hca},
which sets the gluino mass to $m_{\tilde{g}} = 607.71$ GeV. A few example Feynman diagrams are shown in figure \ref{fig:feyndiag}.

This process presents a four-fold permutation symmetry in the final state,
allowing us to explore techniques that can incorporate permutation invariance in
the generative model. Furthermore, the phase space is eight-dimensional, which
conveniently divides into four sets of two variables for every gluino, making
permutation of the phase space straightforward. We choose to use the polar and
azimuthal angles of the gluinos in the center-of-mass frame as parameterization.
Note that this forces the normalizing flow to learn a nontrivial distribution
due to phase space alone, as the phase space measure vanishes in some
regions.\footnote{For instance, it is not possible for all gluino momenta to lie
in the same hemisphere.} The polar and azimuthal angle are mapped to a space
$\mathcal{X} = [0,1]^8$ through
\begin{equation}
    x_{\theta,i} = \frac{1}{2} \left( \cos\theta_i + 1 \right) \text{ and } x_{\varphi_i,i} = \frac{\varphi}{2\pi} \text{ for } i=1,...,4 \,.
\end{equation}

The process shown in eq.~\eqref{eq:process} also presents a rich discrete
structure, which enables an examination of techniques that model continuous and
discrete features simultaneously. Six objects in the adjoint representation of
$\text{SU}(3)_{\text{c}}$ lead to 120 leading-$N_{\text{c}}$ colour-orderings,
and the gluino masses produce a varied spectrum in the 64 helicity
configurations.

Finally, for experiments with varying dimensionality, we mix in $g \kern 0.05em
g \rightarrow \tilde{g} \kern 0.05em \tilde{g}$ events. In total, we generate
$1.2$M $g \kern 0.05em g \rightarrow \tilde{g} \kern 0.05em \tilde{g} \kern
0.05em \tilde{g} \kern 0.05em \tilde{g}$ and $120$k $g \kern 0.05em g\rightarrow
\tilde{g} \kern 0.05em \tilde{g}$ events, reserving $100$k and $10$k for both
validation and testing respectively. The two-gluino events are parameterized by
their common polar and azimuthal angles in the center-of-mass frame.

\subsection{Baseline Normalizing Flow}
We employ a baseline normalizing flow architecture to learn continuous densities
throughout the following experiments. We choose to make use of an autoregressive
flow \cite{papamakarios2017masked} similar to the one used in
\cite{Stienen:2020gns,Caron:2021wmq}. In this model, the bijection $f$ on a
$d$-dimensional event space $\mathcal{X}$ is factorized into a set of $d$
one-dimensional transforms characterized by 
\begin{equation} \label{eq:autoregressive-transform}
    x_j = f_j(z_j; \theta_j(z_{1:j-1})),
\end{equation}
where for $j \in [1,d]$, $z_j$ is the $j$th component of $z$. The bijection
$f_j$ is thus parameterized by a function $\theta_j$ of the preceding
components $z_0$ through $z_{j-1}$. As such, the forward transform from $z$ to
$x$ must be performed sequentially starting from $z_0$. On the contrary, the
inverse transform from $x$ to $z$ can be performed in parallel. This choice
means that training and inference is fast, but sampling is relatively slow
\footnote{Sampling events on a GPU is still fast, taking approximately $20$
seconds for $10^6$ events in our experiments.}. In some of the following
experiments, the sampling step of a normalizing flow is instead required during
training. In such cases, the architecture is inverted such that that direction
is fast. In our implementation, the functional form of $f_j$ is given by a
rational quadratic spline \cite{durkan2019neural} and $\theta_j$ is a MADE
network \cite{germain2015made}. These spline transforms are easily
constrained to a finite domain, making them well-suited for density estimation
in particle physics as phase space can usually be mapped to a finite volume.

In several cases, conditioning of the normalizing flow on some discrete value is
required. That is, instead of just modelling a density $p(x)$ over the
continuous space $\mathcal{X}$, the flow needs to represent a density $p(x|y)$,
where $y \in \mathcal{Y}$ is a discrete number. This type of conditioning
proceeds through learnable embeddings of the values of $y$ into a continuous
space of the size of the hidden layers of the MADE network. These embeddings are
then added before the first activation of the MADE network of every flow layer.

The normalizing flow and all extensions to it discussed in this section are
implemented in \textsc{PyTorch} \cite{paszke2019pytorch}. The code is publically
available\footnote{\url{https://github.com/rbvh/surflows}}. The hyperparameters of the
flow are listed in table \ref{tab:gluino-hyper}. The base distribution $p(z_0)$
is chosen to be a uniform distribution over $[0,1]^8$, such that the flow
transforms are constrained to $[0,1] \to [0,1]$. Models are trained with the
Adam optimizer \cite{kingma2014adam} with default values of $\beta_1$ and
$\beta_2$. Because some of our experiments feature different amounts of training
data, we formulate the training procedure, of which the parameters are also
listed in table \ref{tab:gluino-hyper}\footnote{We find that large batch sizes
lead to better performance. A batch size of $25$k requires $\sim 5$ GB of VRAM,
which is readily available on most modern GPUs.}, in terms of iterations rather
than epochs. After fixed intervals, the model is validated against the
validation set. If the loss has not improved for a fixed number of validations
(the decay patience), the learning rate is multiplied by the decay factor. This
procedure repeats until the learning rate drops low enough for training to have
effectively ceased (in practice, a factor of $10^{-3}$ of the initial learning
rate), or until $5000$ validations have occurred. The model is then finally
evaluated on the independent test set.

We emphasize that the experiments performed in this work are not focussed on
obtaining the best possible performance of the baseline normalizing flow.
Instead, their point is to explore various techniques that one can use to
improve performance on the types of data that are not easily modelled by a
regular normalizing flow, but that regularly appear in the context of particle
physics. Previous work
\cite{Diefenbacher:2020rna,Winterhalder:2021ave,Butter:2021csz} has explored
improving the performance of normalizing flows through the application of an
auxiliary classifier neural network, which can in principle be applied in the
experiments that follow. 

\begin{table}[]
\centering
\begin{tabular}{ll|ll}
\multicolumn{2}{c|}{Model}             & \multicolumn{2}{c}{Training} \\ \hline
Parameter           & Value     & Parameter             & Value       \\ \hline
RQS knots           & 32        & Batch size            & 25k        \\
MADE layers         & 2         & Optimizer             & Adam        \\
MADE units per dim  & 10        & Learning rate         & $10^{-3}$   \\
Flow layers         & 8         & Validation interval   & 25          \\
                    &           & LR decay              & 0.5         \\
                    &           & LR decay patience     & 50          \\
\end{tabular}
\caption{Table of hyperparameters and training setup used in the experiments of section \ref{sec:gluino}.}
\label{tab:gluino-hyper}
\end{table}

\subsection{Permutation Invariance} \label{sec:perm-invariance}
Particle physics events often display a large degree of permutation invariance.
In the matrix element-level example used here, the final state has a four-fold
permutation symmetry. More generally, jet constituents are permutation
invariant, a fact that is already exploited in other ML architectures
\cite{Komiske:2018cqr,Dolan:2020qkr}. Permutation invariance of identified objects also appears at the detector level.

In \cite{nielsen2020survae}, two methods were proposed to instill permutation
invariance into a normalizing flow model: a sorting surjection and a stochastic
permutation. The forward and backward transforms are defined as 
\begin{alignat}{3}
    p_{\text{sort}}(x|z) &= \sum_{\mathcal{I}_p}  \, \frac{1}{D!} \, \delta(x - z_{\mathcal{I}_{p}^{-1}}), && p_{\text{stoch}}(x|z) && = \sum_{\mathcal{I}_p} \, \frac{1}{D!} \, \delta(x - z_{\mathcal{I}_p^{-1}}), \nonumber \\
    q_{\text{sort}}(z|x) &= \sum_{\mathcal{I}_p} \, \delta_{\mathcal{I}_p, \text{argsort}(x)} \, \delta(z - x_{\mathcal{I}_p}),  \quad \quad   && q_{\text{stoch}}(z|x) &&= \sum_{\mathcal{I}_p} \, \frac{1}{D!} \, \delta(z - x_{\mathcal{I}_p}), \nonumber \\
    \mathcal{V}_{\text{sort}}(x,z) &= \log(D!), && \mathcal{V}_{\text{stoch}}(x,z) && =  0,
\end{alignat} \label{eq:permute-transforms}

where $\mathcal{I}_p$ is a set of permutation indices for the components of $x$
or $z$, $\mathcal{I}_p^{-1}$ are their inverse and $D$ is the number of
permutable classes. That is, in the inverse direction, the sort surjection
orders $x$ following some predicate, while the stochastic permutation randomly
shuffles $x$. In the forward direction, both transforms randomly shuffle $z$,
leading to permutation-invariant samples. The sort transform is surjective in
the inverse direction and adheres to the property described in section
\ref{sec:surflows} required for $\mathcal{E}_{\text{sort}}(x,y) = 0$. On the
other hand, stochastic permutation does not lead to a vanishing bound looseness.
Both transforms can lead to improved modelling in different ways. The stochastic
permutation may be viewed as effectively increasing the training statistics by a
factor $D!$, while the sort surjection can be thought of as folding the space
$\mathcal{X}$ into a volume that is a factor $1/D!$ smaller. 

\subsubsection{Experiments}

\begin{figure}
    \centering
    \begin{tikzpicture}
        % Sort figure

        % Nodes
        \node at (-1.7,0.0)(pz0) {$\text{Unif(0,1)}^{d_x} \sim $};
        \node[blob] at (0.0,0.0)(z0) {$z_0$};
        \node[blob] at (2.0,0.0)(z1) {$z_1$};
        \node[blob] at (4.0,0.0)(z2) {$z_2$};
        \node[blob] at (6.0,0.0)(xI) {$x_{\scaleto{\mathcal{I}_{p}}{6pt}}$};
        \node[blob] at (8.0,0.0)(x) {$x$};
            
        % Arrows between zs
        \draw[arrow] (z0) to [out=45,in=135] (z1);
        \draw[arrow] (z1) to [out=225,in=315] (z0);
        \draw[arrow] (z1) to [out=45,in=135] (z2);
        \draw[arrow] (z2) to [out=225,in=315] (z1);
        \draw[arrow] (z2) to [out=45,in=135] (xI);
        \draw[arrow] (xI) to [out=225,in=315] (z2);
        \draw[dashed,arrow] (xI) to [out=45,in=135] (x);
        \draw[arrow] (x) to [out=225,in=315] (xI);
        
        \fill[white](4.9,0.5) rectangle(5.1,0.7);
        \fill[white](4.9,-0.5) rectangle(5.1,-0.7);
        \node at (5.0,0.6) {\footnotesize $\sslash$};
        \node at (5.0,-0.6) {\footnotesize $\sslash$};

        % Function transforms
        \node at (1.0, 0.0) {$f_1$};
        \node at (3.0, 0.0) {$f_2$};
        \node at (5.,0) {...};
        \node at (7.0, 0.0) {sort};       
        
        % Stoch figure

        % Nodes
        \node at (-1.7,-2.0)(pz0) {$\text{Unif(0,1)}^{d_x} \sim $};
        \node[blob] at (0.0,-2.0)(z0) {$z_0$};
        \node[blob] at (2.0,-2.0)(z1) {$z_1$};
        \node[blob] at (4.0,-2.0)(z2) {$z_2$};
        \node[blob] at (6.0,-2.0)(xI) {$x_{\scaleto{\mathcal{I}_{p}}{6pt}}$};
        \node[blob] at (8.0,-2.0)(x) {$x$};
    
        % Arrows between zs
        \draw[arrow] (z0) to [out=45,in=135] (z1);
        \draw[arrow] (z1) to [out=225,in=315] (z0);
        \draw[arrow] (z1) to [out=45,in=135] (z2);
        \draw[arrow] (z2) to [out=225,in=315] (z1);
        \draw[arrow] (z2) to [out=45,in=135] (xI);
        \draw[arrow] (xI) to [out=225,in=315] (z2);
        \draw[dashed,arrow] (xI) to [out=45,in=135] (x);
        \draw[dashed,arrow] (x) to [out=225,in=315] (xI);

        \fill[white](4.9,-1.5) rectangle(5.1,-1.3);
        \fill[white](4.9,-2.5) rectangle(5.1,-2.7);
        \node at (5.0,-1.4) {\footnotesize $\sslash$};
        \node at (5.0,-2.6) {\footnotesize $\sslash$};

        % Function transforms
        \node at (1.0, -2.0) {$f_1$};
        \node at (3.0, -2.0) {$f_2$};
        \node at (5.,-2.0) {...};
        \node at (7.0, -2.0) {stoch};
    \end{tikzpicture}
    \caption{ The normalizing flow architecture of figure
    \ref{fig:flow-illustration} including a sort surjection or a stochastic
    permutation transform at the end. Solid arrows indicate deterministic
    transform directions, while dashed arrows are stochastic. The base
    distributions have been specified to a multivariate uniform.}
    \label{fig:perm-illustration}
\end{figure}
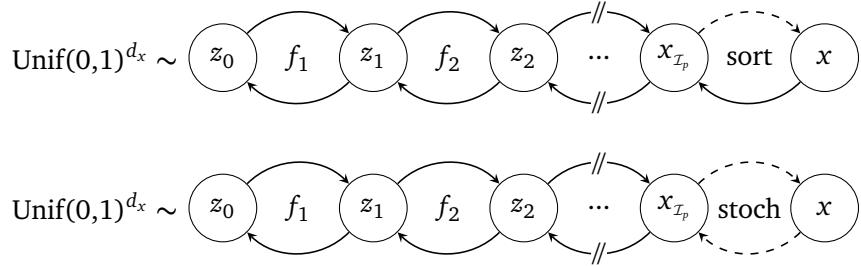
We perform experiments with the default flow model as discussed in the beginning
of this section, either without permutation transform, or with a stochastic
permutation transform, or a sort surjection appended at the end. In this case,
the sort surjection orders gluinos according to their polar angle, as these are
features that are directly present in the phase space parameterization. An
illustration of this architecture is shown in figure
\ref{fig:perm-illustration}.

\begin{figure}
    \includegraphics[width=1.0\textwidth]{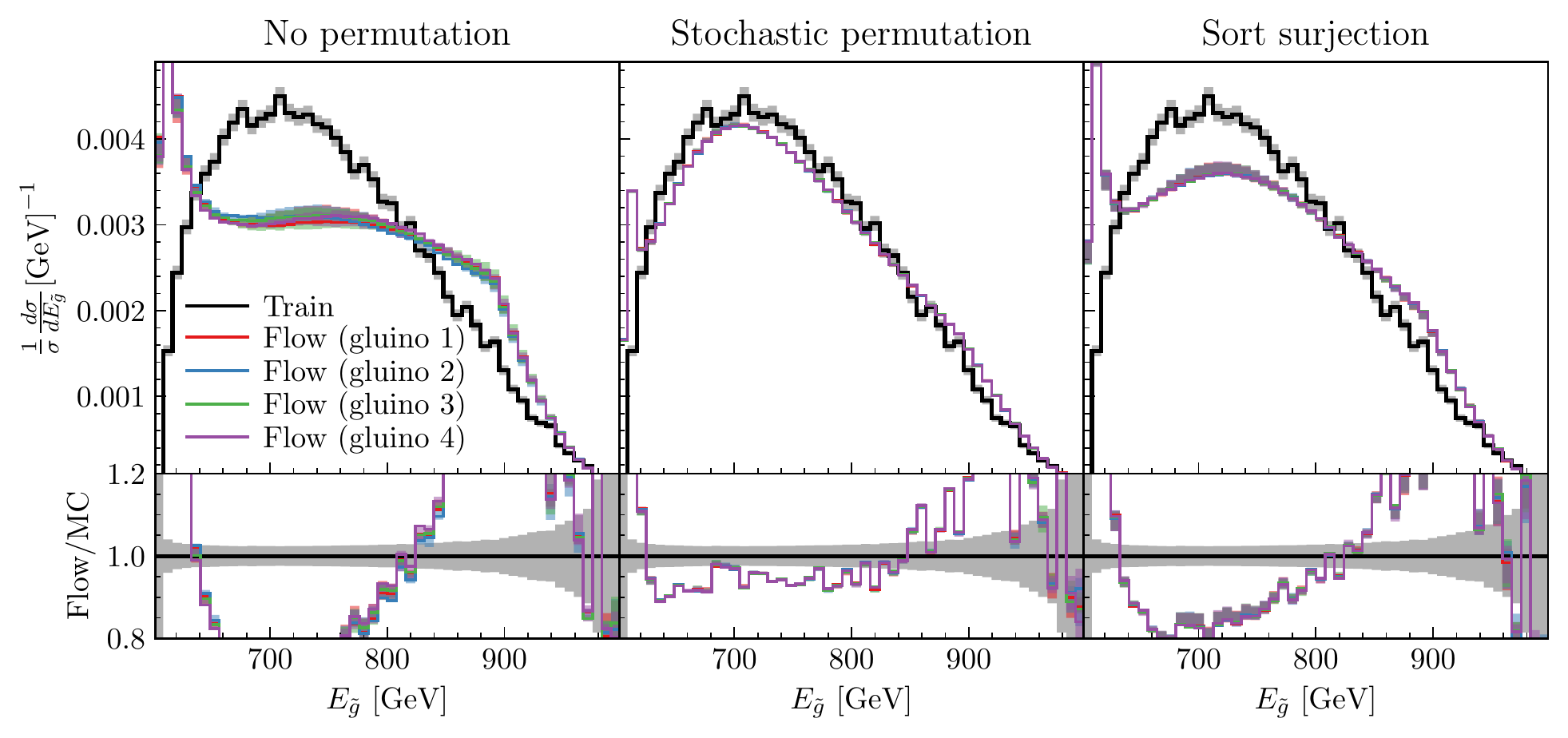}
    \caption{ The energy distribution per individual gluino as predicted by
        models trained on $50$k training events without permutation transform
        (left), with stochastic permutation (middle) or with sort surjection
        (right). The error bands correspond with variations between three
        independent runs.}
    \label{fig:perm-energy}
\end{figure}

To illustrate the gain in performance due to the addition of a permutation
transform, we perform experiments with a varying size of the training dataset.
Figure \ref{fig:perm-energy} shows the distributions of the energy spectra of
the individual gluinos sampled from models trained on just $50$k events. Note
that the gluino energy is not one of the variables that is directly present in
the parameterization of phase space. This means that the model must learn the
relevant correlations between all polar and azimuthal angles to correctly
predict the spectrum. 

We observe that both permutation transforms, and especially the stochastic
permutation, lead to significant improvement in the fidelity of the modeling of
the true distribution. At such small training statistics, the effective increase
with a factor of $4! = 24$ due to the four-fold permutation symmetry is
substantial. Even without permutation transform, the flow mostly learns to treat
the gluinos on equal footing, as only small deviations between the gluino energy
spectra appear. On the other hand, both permutation transforms enforce
permutation invariance in the generative direction, leading to spectra that are
identical up to statistical fluctuations.

\begin{figure}
    \includegraphics[width=1.0\textwidth]{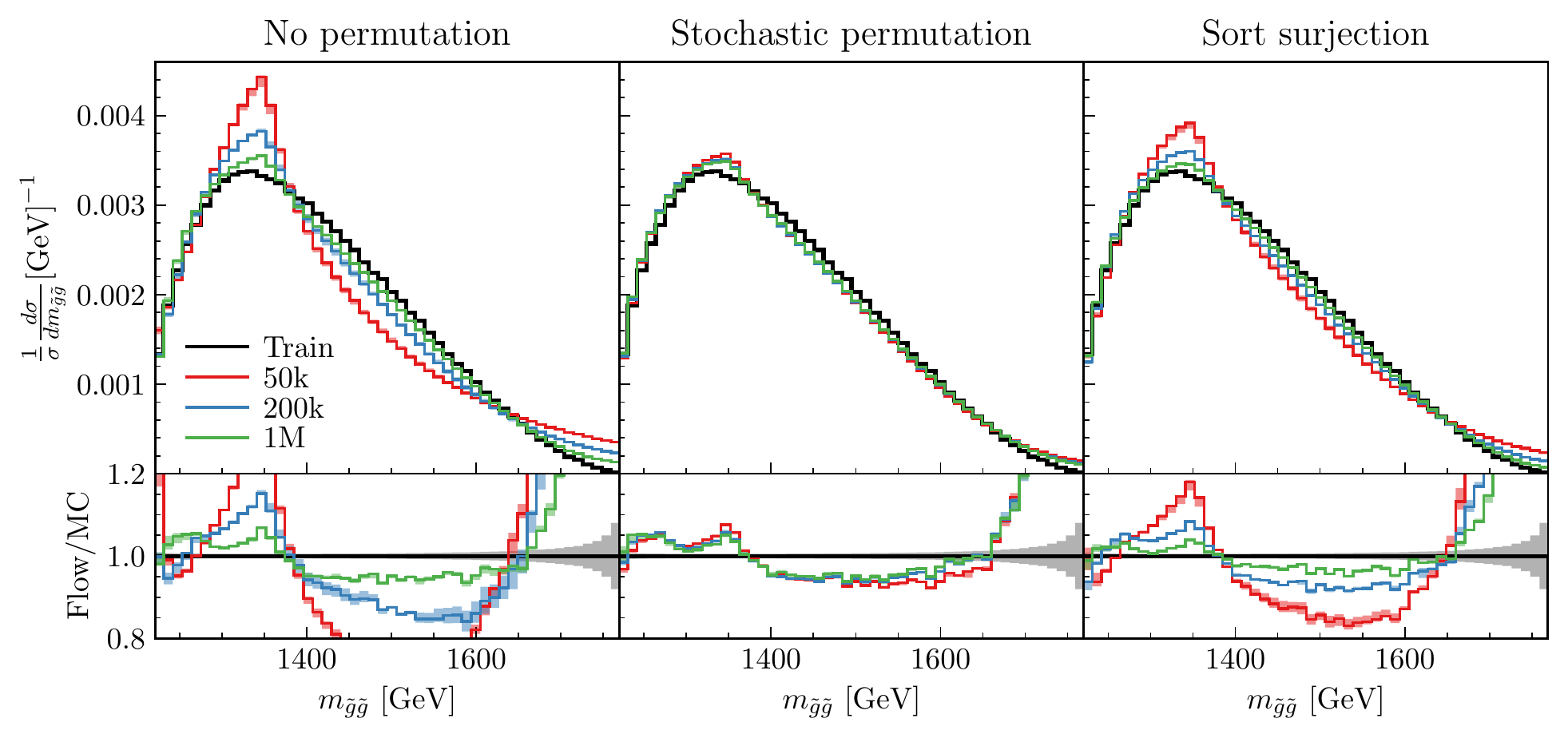}
    \caption{ The digluino invariant mass as predicted by models trained on sets
        of training data of size $50$k (red), $200$k (blue) and $1$M (green),
        without permutation transform (left), with stochastic permutation
        (middle) or with sort surjection (right). The error bands correspond
        with variations between three independent runs.}
    \label{fig:perm-mass}
\end{figure}

\begin{figure}
    \centering
    \includegraphics[width=1.\textwidth]{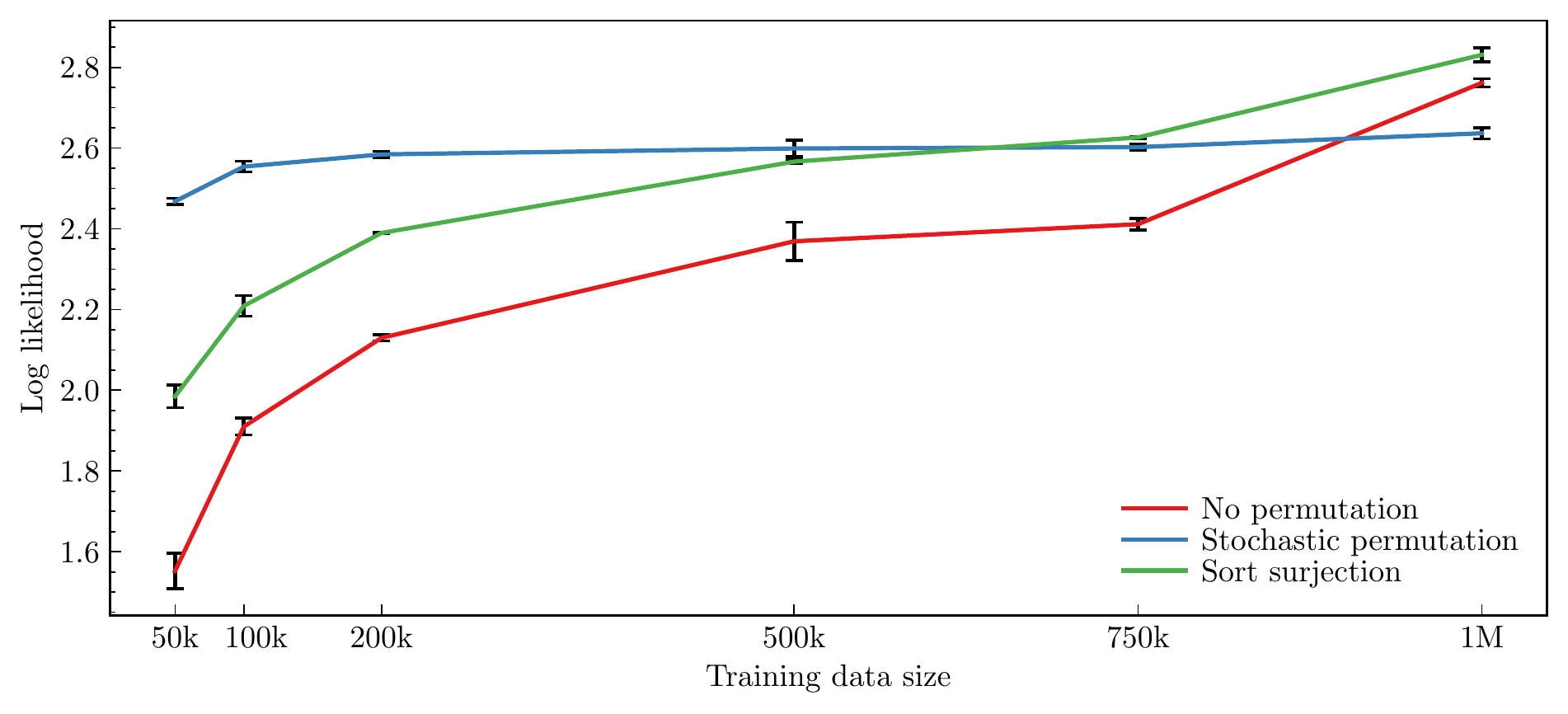}
    \caption{ The development of the test log likelihood (higher is better) of
        models without permutation transform (red), with stochastic permutation
        (blue) or with sort surjection (green). The points and error bars
        correspond with the mean and standard deviation over three independent
        training runs.}
    \label{fig:perm-likelihood}
\end{figure}

Figure \ref{fig:perm-mass} instead shows the digluino invariant mass spectrum,
but this time models trained on $50$k, $200$k and $1$M events are included. One
striking feature of this figure is the fact that the cases without permutation
transform and with sort surjection show definite improvement as the size of the
training dataset increases. However, the case of the stochastic permutation
shows little improvement. This picture is corroborated when one considers the
progression of the testing log likelihood as a function of the size of the
training dataset, which is shown in figure \ref{fig:perm-likelihood}. The model
with stochastic permutation significantly outperforms the other models for small
training statistics, but it is eventually overtaken, even by the model without
permutation transform. This effect occurs due to the nonvanishing bound looseness
associated with the stochastic permutation transform. This means that, given
unconstrained training data and network capacity, the other two cases will
eventually approach the theoretical maximum log likelihood. On the other hand,
the model with stochastic permutation is always limited by a nonzero bound
looseness, diminishing its performance. We conclude that the inclusion of a
permutation transform is always beneficial, but the choice between the two
options should be guided be the size of the available training dataset.

\subsection{Varying Dimensionality} \label{sec:varying-dim}
Particle physics events typically do not contain a constant number of objects.
As a result, the dimensionality of phase space can vary on an event-by-event
basis. Normalizing flow models on the other hand learn probability distributions
of fixed dimension. One method of modelling varying dimensionalities was
presented in \cite{Butter:2021csz} for the specific case of $pp \to Z_{\mu \mu}
+ \{1,2,3\} \text{ jets}$, where conditional flow networks are trained to add
jets to baseline $Z_{\mu \mu}$ events. Alternatively, one could train multiple
generative models for all individual configurations. The downside of this
approach is that the training statistics are split between the models. On the
other hand, a single model that is able to generate all configurations will be
able to learn any underlying patterns that are common between them. The
architecture of \cite{Butter:2021csz} accomplishes this, but it does not
generalize easily to many configurations.

Here, we introduce a surjective transform that is able to combine an arbitrary
number of configurations into a single model.\footnote{The transform introduced
here bears resemblance to the tensor slicing surjection of
\cite{nielsen2020survae}. However, in this case the sliced dimensions are
selected stochastically.} We refer to it as a \emph{dropout} transform, as its
function is to stochastically drop a subset of the latent variables. To that
end, we introduce a set of dropout indices $\mathcal{I}_{\downarrow}$ for the
components of $x$ and $z$, as well as their complement $\mathcal{I}_{\uparrow}$
such that $\{\mathcal{I}_{\downarrow}, \mathcal{I}_{\uparrow}\} = \{1,...,d\}$.
The forward and backward transforms are 
\begin{align} \label{eq:dropout-transform}
    p_{\text{drop}}(x|z) &= \sum_{\mathcal{I}_{\downarrow}} p_{\mathcal{I}_{\downarrow}} \, \delta(x_{\mathcal{I}_{\uparrow}} - z_{\mathcal{I}_{\uparrow}}), \nonumber \\
    q_{\text{drop}}(z|x) &= \sum_{\mathcal{I}_{\downarrow}} \delta_{\mathcal{I}_{\downarrow}, \text{argdrop}(x)} \, \delta(z_{\mathcal{I}_{\uparrow}} - x_{\mathcal{I}_{\uparrow}} ) \, q(z_{\mathcal{I}_{\downarrow}}).
\end{align}
That is, the forward transform picks a set of dropout indices
$\mathcal{I}_{\downarrow}$ with probability $p_{\mathcal{I}_{\downarrow}}$ and
drops the corresponding components from the feature vector $z$. The inverse
fills the dropped components with probability $q(z_{\mathcal{I}_{\downarrow}})$.

The optimal values of the probabilities $p_{\mathcal{I}_{\downarrow}}$ are the
normalized cross-sections of events with configuration
$\mathcal{I}_{\downarrow}$, which can easily be extracted from the training
data, i.e. $p_{\mathcal{I}_{\downarrow}} \equiv
(p_{\mathcal{I}_{\downarrow}})_{\text{data}}$. Note that there is a potential
interaction with the permutation transform of the previous section. When a sort
surjection is used, every feature-space configuration maps to a single set of
dropout indices. However, if a stochastic permutation transform is used, each
feature-space configuration can map to multiple sets of dropout indices, and the
probabilities $p_{\mathcal{I}_{\downarrow}}$ should be adjusted accordingly.

In practice, this means that separate samples of $x$ can have different
dimensionalities. Any following flow layers expect input of the original
dimension of $z$. We handle this by setting dropped indices to values outside
the domain of the relevant latent space.\footnote{For example, in the
experiments performed in this section the latent space is restricted to $[0,1]$,
and dropped indices are set to $-1$.} Subsequent flow layers are then set up to
leave dropped dimensions unchanged. The baseline normalizing flow is conditioned
on the dropout indices.

\begin{figure}
    \centering
    \begin{tikzpicture}
        % Sort figure

        % Nodes
        \node at (-3.75,0.3)(pz0) {$\text{Unif(0,1)}^{d_x} \sim $};
        \node at (-3.1,-0.35)(pcat) {$p_{\mathcal{I}_{\downarrow}} \sim$};
        \node[oval,inner sep=-2pt] at (-2.0,0.0)(z0) {\begin{tabular}{c} $z_0$ \\ $\scaleto{\mathcal{I}_{\downarrow}}{9pt}$ \end{tabular}};
        \node[blob] at (0.0,0.0)(z0drop) {$z_{0, \scaleto{\mathcal{I}_{\uparrow}}{6pt} }$};
        \node[blob] at (2.0,0.0)(z1) {$z_1$};
        \node[blob] at (4.0,0.0)(z2) {$z_2$};
        \node[blob] at (6.0,0.0)(xI) {$x_{\scaleto{\mathcal{I}_{p}}{6pt}}$};
        \node[blob] at (8.0,0.0)(x) {$x$};
            
        % Arrows between zs
        \draw[arrow] (z0) to [out=45,in=135] (z0drop);
        \draw[dashed,arrow] (z0drop) to [out=225,in=315] (z0);
        \draw[arrow] (z0drop) to [out=45,in=135] (z1);
        \draw[arrow] (z1) to [out=225,in=315] (z0drop);
        \draw[arrow] (z1) to [out=45,in=135] (z2);
        \draw[arrow] (z2) to [out=225,in=315] (z1);
        \draw[arrow] (z2) to [out=45,in=135] (xI);
        \draw[arrow] (xI) to [out=225,in=315] (z2);
        \draw[dashed,arrow] (xI) to [out=45,in=135] (x);
        \draw[dashed,arrow] (x) to [out=225,in=315] (xI);
        
        \fill[white](4.9,0.5) rectangle(5.1,0.7);
        \fill[white](4.9,-0.5) rectangle(5.1,-0.7);
        \node at (5.0,0.6) {\footnotesize $\sslash$};
        \node at (5.0,-0.6) {\footnotesize $\sslash$};

        % Function transforms
        \node at (-0.95, 0.0) {drop};
        \node at (1.0, 0.0) {$f_1$};
        \node at (3.0, 0.0) {$f_2$};
        \node at (5.,0) {...};
        \node at (7.0, 0.0) {perm};       
        
    \end{tikzpicture}
    \caption{ The dropout architecture described in section
    \ref{sec:varying-dim}. The permutation transform can either be a stochastic
    permutation or a sort surjection, in which case the arrow in the inverse
    direction would be solid. }
    \label{fig:dropout-illustration}
\end{figure}
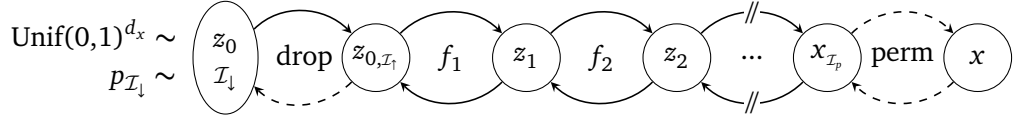

Instead of computing the likelihood contribution and the bound looseness, we can
directly evaluate the marginal likelihood as 
\begin{align} \label{eq:dropout-likelihood}
    p(x) &= \int_{\mathcal{Z}} dz \, p_{\text{drop}}(x|z) \, p(z) \nonumber \\
    &= \sum_{\mathcal{I}_{\downarrow}} p_{\mathcal{I}_{\downarrow}} \int_{\mathcal{Z}_{\mathcal{I}_{\downarrow}}} dz_{\mathcal{I}_{\downarrow}} \, p(x_{\mathcal{I}_{\uparrow}}, z_{\mathcal{I}_{\downarrow}}) \text{ , where } x_{\mathcal{I}_{\uparrow}} = z_{\mathcal{I}_{\uparrow}}.
\end{align} 
In this expression, the latent variables in $p(z)$ have been separated
explicitly into the dropped variables $z_{\mathcal{I}_{\downarrow}}$ and the
remaining ones $z_{\mathcal{I}_{\uparrow}} = x_{\mathcal{I}_{\uparrow}}$. In general, the integral in
eq.~\eqref{eq:dropout-likelihood} is intractable. However, if the distributions
of $z_{\mathcal{I}_{\uparrow}}$ and $z_{I_{\downarrow}}$ are independent, i.e.
$p(x_{\mathcal{I}_{\uparrow}}, z_{\mathcal{I}_{\downarrow}})$ =
$p(x_{\mathcal{I}_{\uparrow}}) \, p(z_{I_{\downarrow}})$,
eq.~\eqref{eq:dropout-likelihood} reduces to 
\begin{equation} \label{eq:dropout-independent}
    p(x) = \sum_{\mathcal{I}_{\downarrow}} \, p_{\mathcal{I}_{\downarrow}} p(x_{\mathcal{I}_{\uparrow}}),
\end{equation}
which can be evaluated exactly. For independence to hold for all
$\mathcal{I}_{\downarrow}$, all latent variables must be mutually independent.
While this is generally not the case after one or more flow layers, the base
distribution is usually chosen as a simple factorized distribution, i.e. a
multivariate uniform or normal distribution with diagonal covariance. Thus, a
model with tractable likelihood emerges by placing the dropout surjection
directly after the base distribution. An illustration of the resulting
architecture is shown in figure \ref{fig:dropout-illustration}. Note that the
stochastic inverse $q(z_{\mathcal{I}_{\downarrow}})$ is no longer required,
since the values of $z_{\mathcal{I}_{\downarrow}}$ are only used to evaluate the
base distribution, and because of the independence from
$z_{\mathcal{I}_{\uparrow}}$ the contribution to the likelihood integrates to
unity.

\subsubsection{Optimization} \label{sec:multi-objective} 
The objective of
 eq.~\eqref{eq:mle} then decomposes into 
\begin{align} \label{eq:mle-dropout}
    \mathcal{L}_{\text{MLE}} &= -\int_{\mathcal{X}} dx \sum_{\mathcal{I}_{\downarrow}} \, p_{\mathcal{I}_{\downarrow}} \, p_{\text{data}}(x_{\mathcal{I}_{\uparrow}}) 
    \log \bigg[ \sum_{\mathcal{I}'_{\downarrow}} \, p_{\mathcal{I}'_{\downarrow}} \, p(x_{\mathcal{I}'_{\uparrow}})  \bigg] \nonumber \\
    &= -\sum_{\mathcal{I}_{\downarrow}} \, p_{\mathcal{I}_{\downarrow}} \, \int_{\mathcal{X}_{\mathcal{I}_{\uparrow}}} dx_{\mathcal{I}_{\uparrow}} \, p_{\text{data}}(x_{\mathcal{I}_{\uparrow}}) 
    \log p(x_{\mathcal{I}_{\uparrow}}) - \underbrace{ \sum_{\mathcal{I}_{\downarrow}} \, p_{\mathcal{I}_{\downarrow}} \log p_{\mathcal{I}_{\downarrow}} }_{\text{constant}}.
\end{align}
That is, maximum likelihood estimation corresponds with a weighted
multi-objective optimization \cite{sener2018multi,zhang2021survey} of the
distributions $p(x_{\mathcal{I}_{\downarrow}})$. Eq.~\eqref{eq:mle-dropout} can
then be interpreted as a linear scalarization of such a multi-objective
optimization problem with preference vector
$p_{\mathcal{I}_{\downarrow}}$. A more general set of solutions to these
problems adhere to the property of Pareto-optimality, which means that one
objective cannot be further improved without degrading at least one of the
others. Linear scalarizations of multi-objective optimization problems like
eq.~\eqref{eq:mle-dropout} can be shown to locate a single Pareto-optimal
solution, but it is often not clear if this is the preferred one. For instance,
when one aims to simultaneously model the distributions of dropout
configurations with widely-varying cross-sections, eq.~\eqref{eq:mle-dropout}
assigns small weight to configurations with small cross-sections, which will
lead to poor modelling of the corresponding conditional probability
distributions. We can instead opt to select a different preference vector
$r_{\mathcal{I}_{\downarrow}}$, leading to 
\begin{equation} \label{eq:mle-dropout-balanced}
    \mathcal{L}_{\text{MLE}}^{(r)} = -\sum_{\mathcal{I}_{\downarrow}} \, r_{\mathcal{I}_{\downarrow}} \, \int_{\mathcal{X}_{\mathcal{I}_{\uparrow}}} dx_{\mathcal{I}_{\uparrow}} \, p_{\text{data}}(x_{\mathcal{I}_{\uparrow}}) 
    \log p(x_{\mathcal{I}_{\uparrow}}).
\end{equation}

\subsubsection{Experiments}
\begin{figure}
    \includegraphics[width=1.\textwidth]{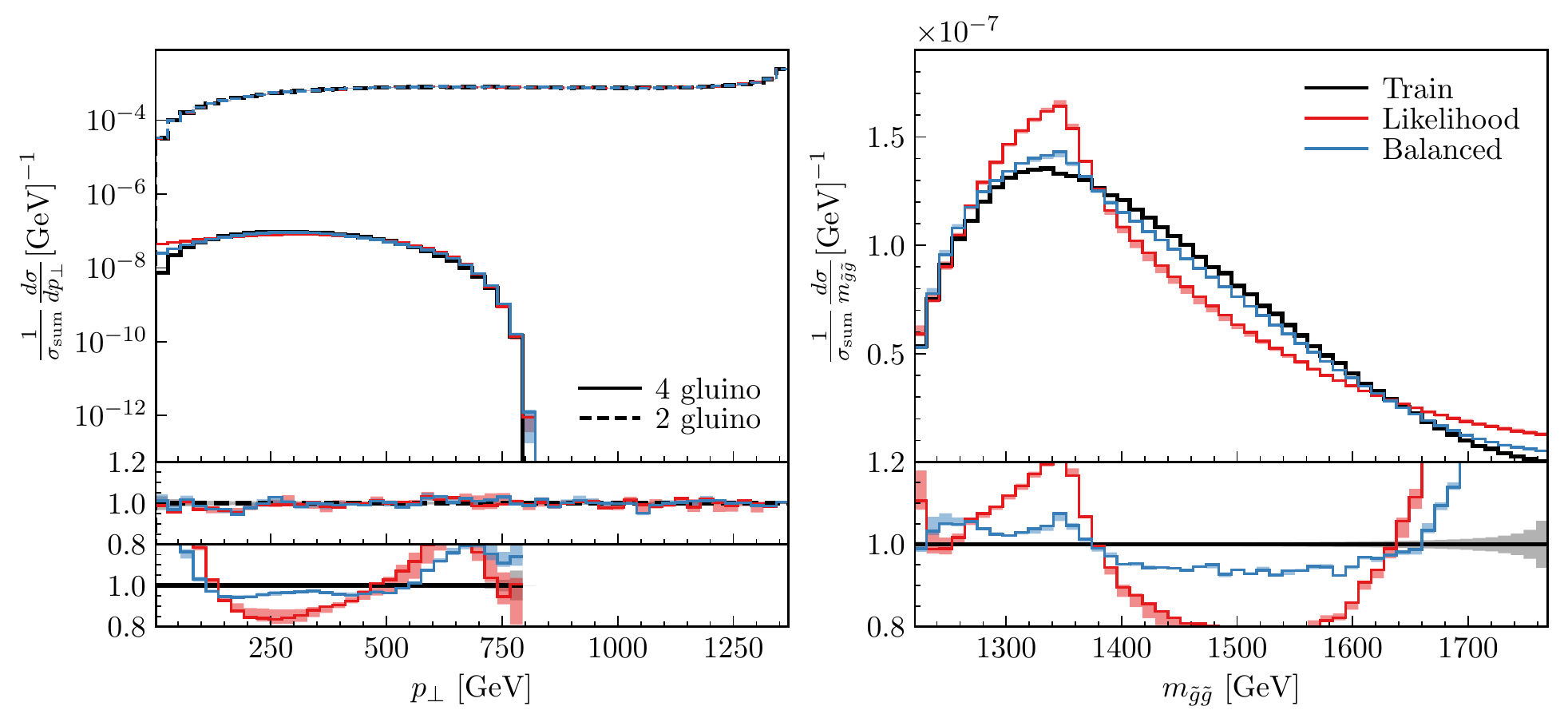}
    \caption{ The transverse-momentum spectrum (left) of four-gluino events
        (solid) and two-gluino events (dashed), as well as the digluino
        invariant mass spectrum (right) of four-gluino events. The MC truth
        (black) is compared with the model illustrated in figure
        \ref{fig:dropout-illustration} trained by maximizing
        eq.~\eqref{eq:mle-dropout} (red) and eq.~\eqref{eq:mle-dropout-balanced}
        with the elements of the weight vector $r_{\mathcal{I}_{\downarrow}}$
        set to $1/2$ (blue). The error bands correspond with variations between
        three independent runs.}
    \label{fig:dropout-spectra}
\end{figure}

We perform experiments with the model of section \ref{sec:perm-invariance} with
ordering surjection, and include a dropout layer directly after the base
distribution. The objective is to learn the distributions of two-gluino and
four-gluino events simultaneously. 
There are only two non-vanishing dropout probabilities
\begin{align}
    p_{\text{two}} &= 4.0069 \cdot 10^{-5} &&\mathcal{I}_{\downarrow} = \{3,4,5,6,7,8\}, \,\, \mathcal{I}_{\uparrow} = \{1,2\} \nonumber \\
    p_{\text{four}} &= 1 -4.0069 \cdot 10^{-5} && \mathcal{I}_{\downarrow} = \{\} , \,\, \mathcal{I}_{\uparrow} = \{1,2,3,4,5,6,7,8\}.
\end{align}
The values of the dropout likelihoods follow the cross-sections of the
corresponding processes. Training is performed by either maximizing
eq.~\eqref{eq:mle-dropout} (referred to as \emph{likelihood}), or by maximizing
eq.~\eqref{eq:mle-dropout-balanced} (referred to as \emph{balanced}) where the
elements of the weight vector are set to $1/2$. The large difference in
cross-section would lead to a very small amount of four-gluino events in the
training data. We instead opt to use the full datasets described in section
\ref{sec:benchmark} and reweigh as appropriate.

Figure \ref{fig:dropout-spectra} shows the spectra of the transverse momentum of
four-gluino and two-gluino events, as well as the digluino invariant mass of the
four-gluino events. We observe significantly better performance in the balanced
case. This is the result of the small weight assigned to the four-gluino
conditional in the likelihood case, resulting in poor optimization. Experiments
with different values of the weight vector $r_{\mathcal{I}_{\downarrow}}$ did not
lead to qualitatively different results, which appears to indicate one can expect
similar performance as long as none of the conditional distributions of
eq.~\eqref{eq:mle-dropout-balanced} are substantially suppressed.

\subsection{Discrete Features} \label{sec:discrete}
Generative models in particle physics have predominantly focussed on modelling
the continuous phase space of particle collisions. However, scattering events
are often not only characterized by their energy-momentum distributions, but
also by a variety of discrete features which are related to the quantum numbers
of the particles involved in the scattering.

Several methods of modelling discrete features have been considered in the
context of normalizing flows. Some of these find explicit methods of casting
eq.~\eqref{eq:flow-step-llh} in a form that can handle discrete data
\cite{tran2019discrete,ziegler2019latent}. However, particle physics presents a
distinct situation in which continuous and discrete features are jointly
distributed, and these methods are not straightforwardly extended to model
correlations between continuous and discrete data components. We thus explore
methods that either map the discrete space to a continuous one, or which
explicitly factorize the two spaces.

We consider a situation where points in the data space may be denoted as
$(x,y) \in (\mathcal{X}, \mathcal{Y})$, where $\mathcal{X}$ is continuous and
$d_x$-dimensional, i.e. $x_i \in [0,1]^{d_{x}}$, and $\mathcal{Y}$ is discrete
and $d_y$-dimensional, i.e. $y_i \in \prod_{i=1}^{d_y} \{0,...,N_i-1\}$.

\subsubsection{Variational Dequantization} \label{sec:deq}
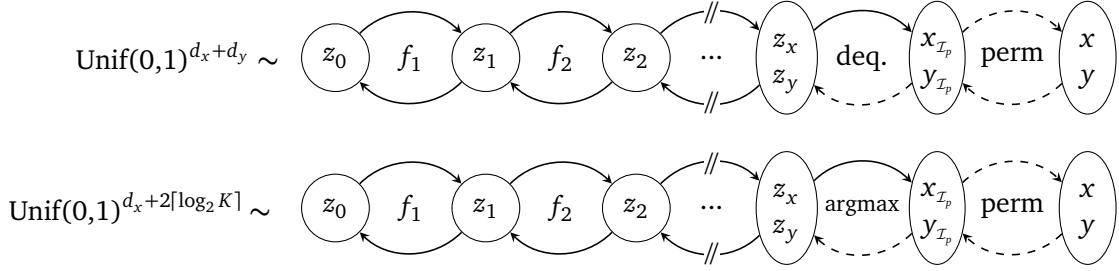
\begin{figure}
    \centering
    \begin{tikzpicture}
        % Dequantization figure

        % Nodes
        \node at (-2.1,0.0)(pz0) {$\text{Unif(0,1)}^{d_x + d_y} \sim $};
        \node[blob] at (0.0,0.0)(z0) {$z_0$};
        \node[blob] at (2.0,0.0)(z1) {$z_1$};
        \node[blob] at (4.0,0.0)(z2) {$z_2$};
        \node[oval,inner sep=-3pt] at (6.0,0.0)(zxy) {\begin{tabular}{c} $z_x$ \\ $z_y$ \end{tabular}};
        \node[oval,inner sep=-5pt] at (8.0,0.0)(xyI) {\begin{tabular}{c} $x_{\scaleto{\mathcal{I}_{p}}{6pt}}$ \\ $y_{\scaleto{\mathcal{I}_{p}}{6pt}}$ \end{tabular}};
        \node[oval,inner sep=-2pt] at (10.0,0.0)(xy) {\begin{tabular}{c} $x$ \\ $y$ \end{tabular}};
            
        % Arrows between zs
        \draw[arrow] (z0) to [out=45,in=135] (z1);
        \draw[arrow] (z1) to [out=225,in=315] (z0);
        \draw[arrow] (z1) to [out=45,in=135] (z2);
        \draw[arrow] (z2) to [out=225,in=315] (z1);
        \draw[arrow] (z2) to [out=45,in=135] (zxy);
        \draw[arrow] (zxy) to [out=225,in=315] (z2);
        \draw[arrow] (zxy) to [out=45,in=135] (xyI);
        \draw[dashed,arrow] (xyI) to [out=225,in=315] (zxy);
        \draw[dashed,arrow] (xyI) to [out=45,in=135] (xy);
        \draw[dashed,arrow] (xy) to [out=225,in=315] (xyI);

        \fill[white](4.9,0.5) rectangle(5.1,0.7);
        \fill[white](4.9,-0.5) rectangle(5.1,-0.7);
        \node at (5.0,0.6) {\footnotesize $\sslash$};
        \node at (5.0,-0.6) {\footnotesize $\sslash$};

        % Function transforms
        \node at (1.0, 0.0) {$f_1$};
        \node at (3.0, 0.0) {$f_2$};
        \node at (5.,0) {...};
        \node at (7.0, 0.0) {deq.};
        \node at (9.0, 0.0) {perm}; 

        % Dequantization figure
        
        % Nodes
        \node at (-2.6,-2.0)(pz0) {$\text{Unif(0,1)}^{d_x + 2\lceil \log_2K \rceil} \sim $};
        \node[blob] at (0.0,-2.0)(z0) {$z_0$};
        \node[blob] at (2.0,-2.0)(z1) {$z_1$};
        \node[blob] at (4.0,-2.0)(z2) {$z_2$};
        \node[oval,inner sep=-3pt] at (6.0,-2.0)(zxy) {\begin{tabular}{c} $z_x$ \\ $z_y$ \end{tabular}};
        \node[oval,inner sep=-5pt] at (8.0,-2.0)(xyI) {\begin{tabular}{c} $x_{\scaleto{\mathcal{I}_{p}}{6pt}}$ \\ $y_{\scaleto{\mathcal{I}_{p}}{6pt}}$ \end{tabular}};
        \node[oval,inner sep=-2pt] at (10.0,-2.0)(xy) {\begin{tabular}{c} $x$ \\ $y$ \end{tabular}};
            
        % Arrows between zs
        \draw[arrow] (z0) to [out=45,in=135] (z1);
        \draw[arrow] (z1) to [out=225,in=315] (z0);
        \draw[arrow] (z1) to [out=45,in=135] (z2);
        \draw[arrow] (z2) to [out=225,in=315] (z1);
        \draw[arrow] (z2) to [out=45,in=135] (zxy);
        \draw[arrow] (zxy) to [out=225,in=315] (z2);
        \draw[arrow] (zxy) to [out=45,in=135] (xyI);
        \draw[dashed,arrow] (xyI) to [out=225,in=315] (zxy);
        \draw[dashed,arrow] (xyI) to [out=45,in=135] (xy);
        \draw[dashed,arrow] (xy) to [out=225,in=315] (xyI);

        \fill[white](4.9,-1.5) rectangle(5.1,-1.3);
        \fill[white](4.9,-2.5) rectangle(5.1,-2.7);
        \node at (5.0,-1.4) {\footnotesize $\sslash$};
        \node at (5.0,-2.6) {\footnotesize $\sslash$};

        % Function transforms
        \node at (1.0, -2.0) {$f_1$};
        \node at (3.0, -2.0) {$f_2$};
        \node at (5.,-2.0) {...};
        \node at (7.0, -2.0) {\footnotesize argmax};
        \node at (9.0, -2.0) {perm}; 
    \end{tikzpicture}
    
    \caption{ The variational dequantization and argmax surjection transform
    architectures described in sections \ref{sec:deq} and \ref{sec:argmax}. The
    permutation transform can either be a stochastic permutation or a sort
    surjection, in which case the arrow in the inverse direction would be solid.
    }
    \label{fig:deq-illustration}
\end{figure}

One option is to introduce a surjective transform that adds noise around the
discrete values in the inverse direction. The result can then be appended to
the other continuous features, and the normalizing flow can be used to learn
correlations between the continuous and the discrete components. This process is
often referred to as \emph{variational dequantization}
\cite{uria2013rnade,salimans2017pixelcnn++,ho2019flow++,nielsen2020closing}. The
forward and backward transforms are \cite{nielsen2020survae}
\begin{align} \label{eq:dequantization}
    p_{\text{deq}}(x,y|z) &= \delta(x - z_x) \, \mathbb{I}_{F(y)}(z_y), \nonumber \\
    q_{\text{deq}}(z|x,y) &= \delta(z_x - x) \, q_{\text{deq.}}(z_y| y), \nonumber \\
    \mathcal{V}_{\text{deq}}(x,y,z) &= -\log q_{\text{deq.}}(z_y| y),
\end{align}
where $\mathbb{I}_{F(y)}$ is the indicator function over the set 
\begin{equation}
    F(y) = \{y + u_y | u_y \in [0,1)^d\}
\end{equation}
and the support of $q_{\text{deq.}}(z_y| y)$ is restricted to $F(y)$. The
continuous latent variables are thus split up into two groups, $z_x = x \in
[0,1]^{d_x}$ and $z_y \in [0,1]^{d_y}$ which get mapped to $y$. The resulting
model is illustrated in figure \ref{fig:deq-illustration}.

Variational dequantization leads to a nonvanishing bound looseness, but a more
flexible implementation of the dequantizing distribution $q_{\text{deq.}}(z_y|
y)$ can reduce its size \cite{hoogeboom2020learning}. In our experiments, we
include a model in which the dequantizer $q_{\text{deq.}}(z_y| y)$ samples
$u_y$ uniformly, and one in which it is sampled by an auxiliary flow model
conditioned on $y$. In the former case, the main flow model is tasked with
learning a discontinuous probability distribution, while in the latter case the
auxiliary flow can populate the disjoint sets $F(y)$ such that the distribution
over $z_y$ is smoother. Both flows are trained jointly during the optimization of
eq.~\eqref{eq:elbo-2}.

\subsubsection{Argmax Surjection} \label{sec:argmax}
Variational dequantization is particularly well-suited for ordinal discrete
data, where adjacent categories are often associated with similar likelihoods.
However, in most situations in particle physics, the discrete data at hand are
(derived from) quantum numbers which are fundamentally categorical.
Dequantization may thus not be the optimal method of treating them. In
\cite{hoogeboom2021argmax} an \emph{argmax surjection} was introduced to handle
categorical data. This transform makes minimal assumptions about the topology of
the discrete features. Individual categories are equidistant, and the continuous
space is evenly partitioned. 

In the context of fully equidistant categories, the dimension $d_y$ of
$\mathcal{Y}$ loses its meaning, and without loss of generality we can instead
consider $y \in \{0,...,K-1\}$, where $K = \prod_{i=1}^{d_y} N_i$. The
transforms of the argmax surjection are also given by
eq.~\eqref{eq:dequantization}, and an illustration of the architecture is shown
in figure \ref{fig:deq-illustration}. However, the supporting set is now given
by 
\begin{equation} \label{eq:argmax-support-set}
    F(y) = \bigg\{z_y \bigg| \argmax_{k\in \{0,...,K-1\}} (z_y)_k = y\bigg\}.
\end{equation}
That is, $z_y \in [0,1]^K$ and the argmax operation selects $k$ if $\forall_{i
\neq k} (z_y)_i < (z_y)_k$, where $(z_y)_i$ is the $i$th component of $z_y$.

A major downside of this approach is the fact that the dimensionality of the
latent space is now $d_x + K$ instead of $d_x + d_y$. In the experiments we
perform here (where $K = 7680$), and in most other situations, a latent space of
this magnitude is not manageable. The authors of \cite{hoogeboom2021argmax}
propose to instead model a binary decomposition $y_{\text{B}} =
\{0,1\}^{\log_2K}$, such that every bit requires only two latent dimensions
corresponding with the values $0$ and $1$. The total required latent space
dimension then reduced to $d_x + 2\lceil \log_2K \rceil$ (34 in our
experiments). Note that this choice essentially represents a compromise between
symmetry between the individual labels, and the dimensionality of the problem.
One could thus implement other decompositions of the categorical space that
would lead to a different balance.

Note that, during sampling, it is now possible for the model to generate binary
encodings that correspond with categorical labels that are outside the range of
the data. Of course, the model should learn to assign very small likelihoods to
such events. If this happens, the complete event is rejected, and a new one is
generated.

As in section \ref{sec:deq}, we perform experiments with a uniform argmax
dequantizer, as well as with an auxiliary flow in an attempt to find the
configuration that minimizes the bound looseness. The dequantizer is restricted
to the supporting set of eq.~\eqref{eq:argmax-support-set} by first sampling
$z'_y \in ([0,1]^2)^{\lceil \log_2K\rceil}$. Then, a variable transform is
performed to restrict the result to the set $F(y)$. For example, if $z_{y,1} >
z_{y,2}$ is required to reproduce a particular bit, the transform is 
\begin{align}
    z_{y,1} &= z'_{y,1} \nonumber \\
    z_{y,2} &= z'_{y,1} z'_{y,2}.
\end{align}
The transform from $z'_{y,1}, z'_{y,2} \to z_{y,1}, z_{y,2}$ induces an
additional Jacobian factor $\log |J| = -\log z'_{y,1}$ that is incorporated in
the evaluation of $q_{\text{deq.}}(z_y| y)$.

\subsubsection{Factorized Models} \label{sec:factorized}
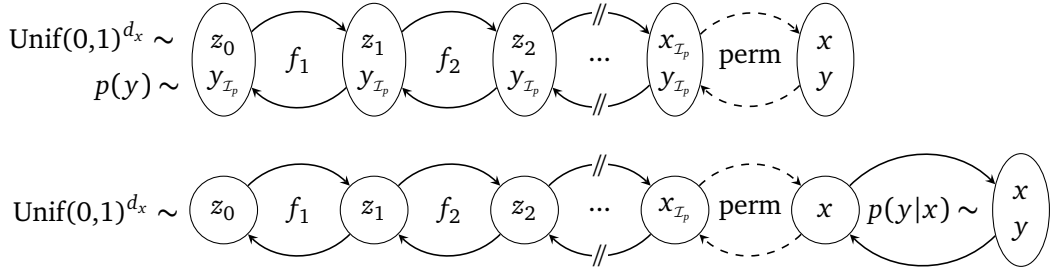
\begin{figure}
    \centering
    \begin{tikzpicture}
        % Sort figure

        % Nodes
        \node at (-1.75,0.3)(pz0) {$\text{Unif(0,1)}^{d_x} \sim $};
        \node at (-1.15,-0.35)(pcat) {$p(y) \sim$};
        \node[oval,inner sep=-4pt] at (0.0,0.0)(z0y) {\begin{tabular}{c} $z_0$ \\ $y_{\scaleto{\mathcal{I}_{p}}{6pt}}$\end{tabular}};
        \node[oval,inner sep=-4pt] at (2.0,0.0)(z1) {\begin{tabular}{c} $z_1$ \\ $y_{\scaleto{\mathcal{I}_{p}}{6pt}}$\end{tabular}};
        \node[oval,inner sep=-4pt] at (4.0,0.0)(z2) {\begin{tabular}{c} $z_2$ \\ $y_{\scaleto{\mathcal{I}_{p}}{6pt}}$\end{tabular}};
        \node[oval,inner sep=-5pt] at (6.0,0.0)(xI) {\begin{tabular}{c} $x_{\scaleto{\mathcal{I}_{p}}{6pt}}$ \\ $y_{\scaleto{\mathcal{I}_{p}}{6pt}}$\end{tabular}};
        \node[oval,inner sep=-2pt] at (8.0,0.0)(x) {\begin{tabular}{c} $x$ \\ $y$\end{tabular}};
            
        % Arrows between zs
        \draw[arrow] (z0y) to [out=45,in=135] (z1);
        \draw[arrow] (z1) to [out=225,in=315] (z0y);
        \draw[arrow] (z1) to [out=45,in=135] (z2);
        \draw[arrow] (z2) to [out=225,in=315] (z1);
        \draw[arrow] (z2) to [out=45,in=135] (xI);
        \draw[arrow] (xI) to [out=225,in=315] (z2);
        \draw[dashed,arrow] (xI) to [out=45,in=135] (x);
        \draw[dashed,arrow] (x) to [out=225,in=315] (xI);
        
        \fill[white](4.9,0.5) rectangle(5.1,0.7);
        \fill[white](4.9,-0.5) rectangle(5.1,-0.7);
        \node at (5.0,0.6) {\footnotesize $\sslash$};
        \node at (5.0,-0.6) {\footnotesize $\sslash$};

        % Function transforms
        \node at (1.0, 0.0) {$f_1$};
        \node at (3.0, 0.0) {$f_2$};
        \node at (5.,0) {...};
        \node at (7.0, 0.0) {perm};       

        % Stoch figure

        % Nodes
        \node at (-1.7,-2.0)(pz0) {$\text{Unif(0,1)}^{d_x} \sim $};
        \node[blob] at (0.0,-2.0)(z0) {$z_0$};
        \node[blob] at (2.0,-2.0)(z1) {$z_1$};
        \node[blob] at (4.0,-2.0)(z2) {$z_2$};
        \node[blob] at (6.0,-2.0)(xI) {$x_{\scaleto{\mathcal{I}_{p}}{6pt}}$};
        \node[blob] at (8.0,-2.0)(x) {$x$};
        \node[oval,inner sep=-2pt] at (10.6,-2.0)(xy) {\begin{tabular}{c} $x$ \\ $y$\end{tabular}};
    
        % Arrows between zs
        \draw[arrow] (z0) to [out=45,in=135] (z1);
        \draw[arrow] (z1) to [out=225,in=315] (z0);
        \draw[arrow] (z1) to [out=45,in=135] (z2);
        \draw[arrow] (z2) to [out=225,in=315] (z1);
        \draw[arrow] (z2) to [out=45,in=135] (xI);
        \draw[arrow] (xI) to [out=225,in=315] (z2);
        \draw[dashed,arrow] (xI) to [out=45,in=135] (x);
        \draw[dashed,arrow] (x) to [out=225,in=315] (xI);
        \draw[arrow] (x) to [out=45,in=135] (xy);
        \draw[arrow] (xy) to [out=225,in=315] (x);

        \fill[white](4.9,-1.5) rectangle(5.1,-1.3);
        \fill[white](4.9,-2.5) rectangle(5.1,-2.7);
        \node at (5.0,-1.4) {\footnotesize $\sslash$};
        \node at (5.0,-2.6) {\footnotesize $\sslash$};

        % Function transforms
        \node at (1.0, -2.0) {$f_1$};
        \node at (3.0, -2.0) {$f_2$};
        \node at (5.,-2.0) {...};
        \node at (7.0, -2.0) {perm};
        \node at (9.3, -2.0) {$p(y|x) \sim $};
        
    \end{tikzpicture}
    \caption{ The factorized architectures described in section
    \ref{sec:factorized}. The permutation transform can either be a stochastic
    permutation or a sort surjection, in which case the arrow in the inverse
    direction would be solid. }
    \label{fig:factorized-ilustration}
\end{figure}

An alternative approach is to explicitly factorize the continuous and discrete
densities, leading to models of which the likelihood can be evaluated exactly.
The first option is to factorize the joint density as
\begin{equation}
    p(x,y) = p(y) \, p(x|y).
\end{equation}
In this case, the categorical distribution $p(y)$ is straightforwardly extracted
from the data. The conditional continuous density $p(x|y)$ can be modelled by a
normalizing flow that is conditioned on the category $y$. 

One advantage of this method is that it is easily combined with that of section
\ref{sec:varying-dim} in case the data contains variable numbers of objects that
have both continuous and discrete features, i.e. particles with a four-vector and
an identity. Furthermore, it is guaranteed that the marginalized categorical
distribution $p(y)$ matches the training data exactly, which is not the case for
the previous methods. We refer to this approach as a \emph{mixture} model.

The second option is to instead factorize the joint density as 
\begin{equation}
    p(x,y) = p(x) \, p(y|x).
\end{equation}
Now, $p(x)$ is the same type of normalizing flow as was considered in section
\ref{sec:perm-invariance}, trained by ignoring the discrete features. The
conditional distribution $p(y|x)$ can be implemented as a neural network that
predicts the categorical probabilities of $y$ given an instance of $x$. This task
is in fact identical to a multi-class classification problem, where a neural
network is trained to optimize the categorical cross-entropy between real and
predicted labels. We thus refer to this model as a \emph{classifier}. Both
factorized model architectures are illustrated in
\ref{fig:factorized-ilustration}.

\subsubsection{Experiments} \label{sec:discrete-experiments}
We perform experiments with all methods described in the previous section, using
the full training dataset. The continuous features of the four-gluino events are
appended by the helicity configuration, which is encoded into a single category
by interpreting it as a binary string, and by the colour ordering, which is
converted into its Lehmer code \cite{Lehmer1960TeachingCT}.

For the variational dequantization models, the helicity and colour labels are
kept separate. For all other models, the labels are combined into one, which is
then decomposed into its binary representation for models with the argmax
surjection. Following the discussion of section \ref{sec:multi-objective}, we
train the mixture model using both the likelihood and balanced prescriptions.
For all cases except the classifier, we train models with both a stochastic
permutation transform and a sorting surjection. The classification model is
instead composed of the best-performing normalizing flow with ordering
surjection trained on the full training dataset from section
\ref{sec:perm-invariance}, and of a classifier consisting of a multilayer
perceptron with $3$ hidden layers of $256$ nodes and ReLU activation functions,
followed by a final softmax activation. It is trained following the same
procedure as the normalizing flow, but with an initial learning rate of
$10^{-5}$.

\begin{figure}
    \includegraphics[width=1.\textwidth]{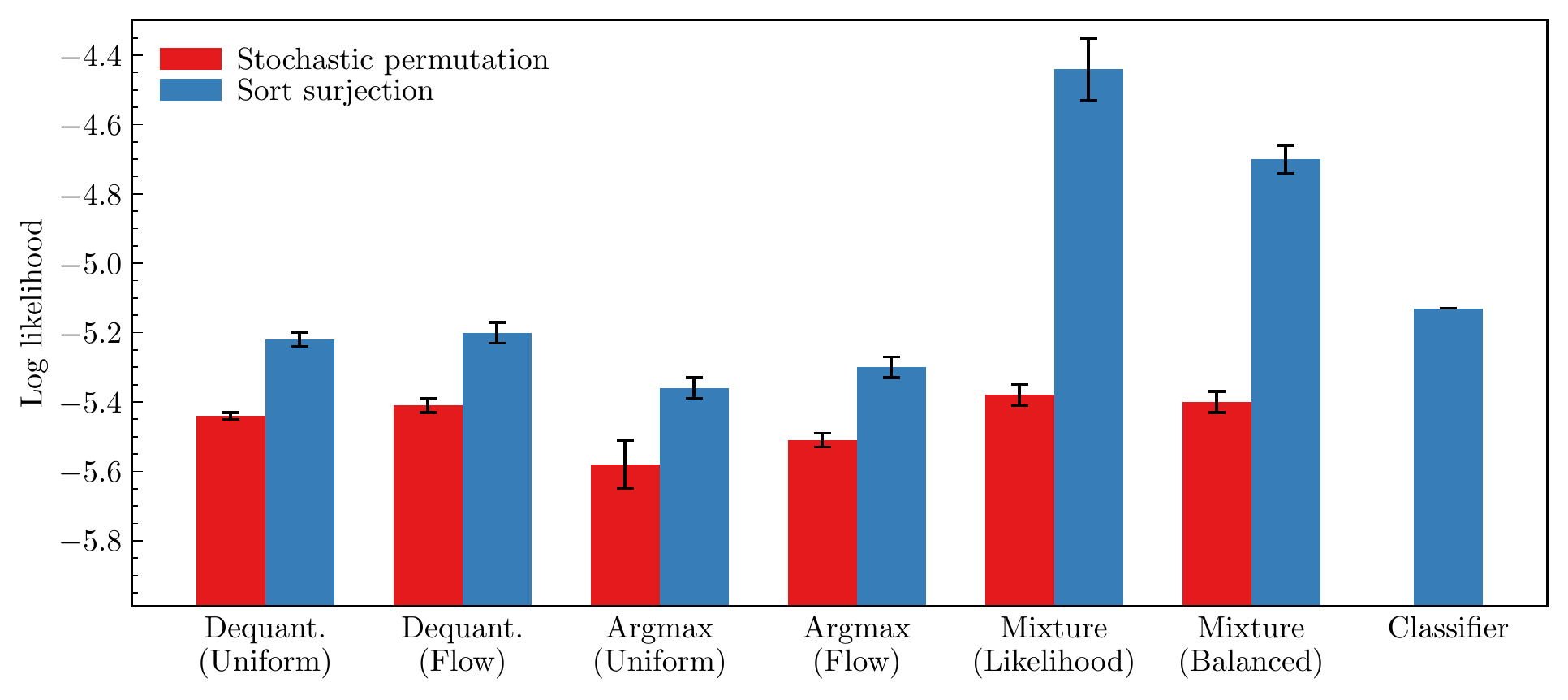}
    \caption{ Test log likelihoods (higher is better) of various generative
        models for density estimation on mixed continuous-discrete data applied
        to four-gluino events, including helicity and colour labels. The bars
        show the average results over three independent runs, and the error bar
        indicates the standard deviation. }
    \label{fig:discrete-losses}
\end{figure}

Figure \ref{fig:discrete-losses} shows the test log likelihood of all models. We
observe especially good performance from the mixture model with sorting
surjection. A likely explanation is that the mixture models benefit from the
fact that the marginalized discrete distributions are correct by construction,
while all other models need to learn them implicitly. The classifier model
performs slightly better than those using variational dequantization or argmax
surjections, which necessarily suffer from nonvanishing bound looseness.
Interestingly, among them, the dequantization options seem to be preferred even
though the discrete features are of categorical nature. The most plausible
explanation is that the normalizing flow is expressive enough to learn the
dequantized distribution of the helicity and colour labels, while the argmax
model suffers from the much larger required latent space. The choice of
populating the supporting sets with a flow instead of a uniform distribution
only leads to marginal improvements, which is again aligned with the observation
that the baseline normalizing flow is expressive enough to learn the
discontinuous distributions presented to it in the uniform case.

\begin{figure}
    \includegraphics[width=1.\textwidth]{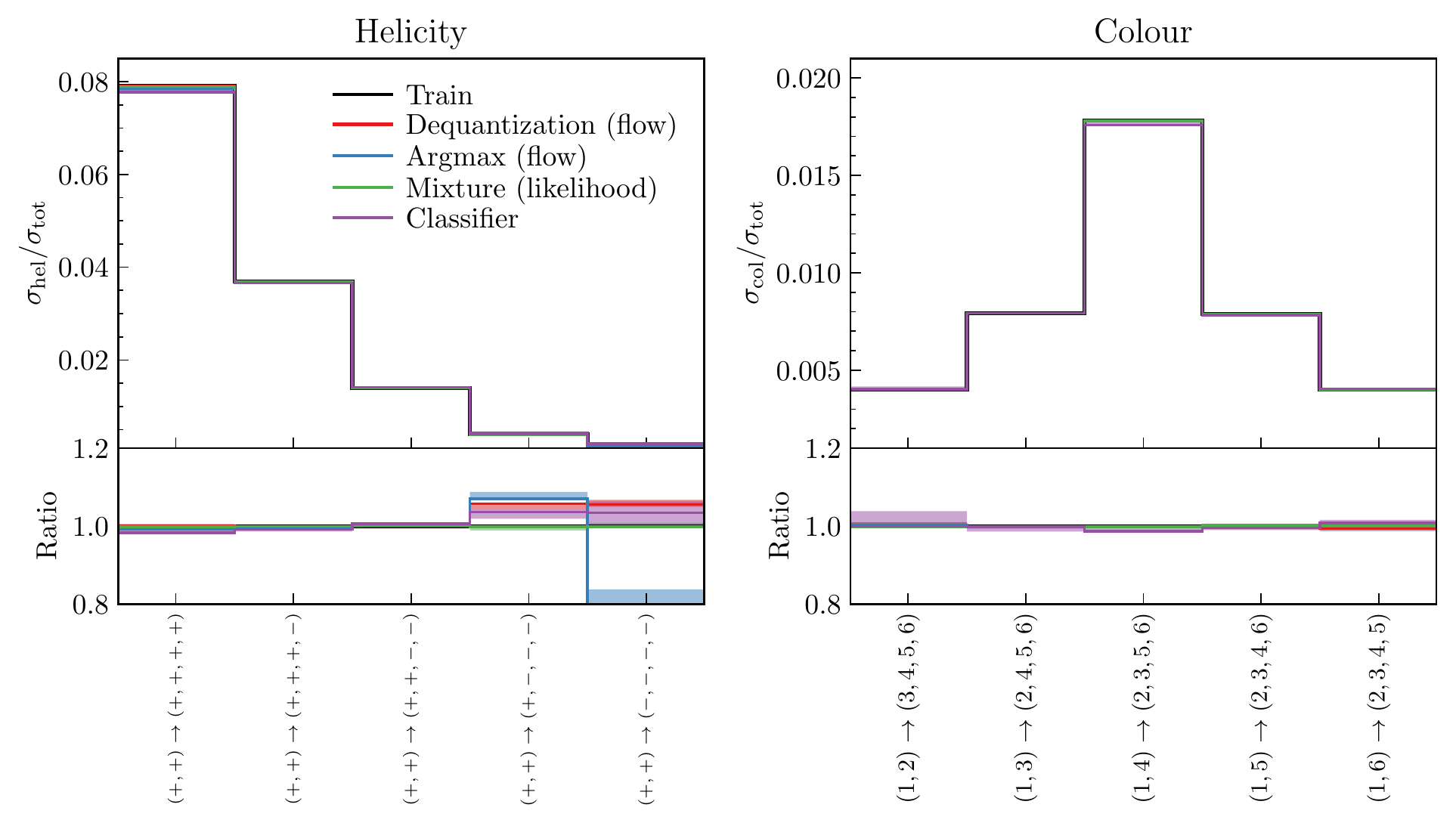}
    \caption{ The marginalized discrete likelihoods of a subset of helicity and
        colour configurations as predicted by models with variational
        dequantization (red), argmax surjection (blue), the mixture model
        (green) and the classifier model (purple). In all cases, a sorting
        surjection was used. The error bands correspond with variations between
        three independent runs.}
    \label{fig:marginal-discrete}
\end{figure}

Figure \ref{fig:marginal-discrete} shows the marginalized discrete likelihoods
of a subset of helicity and colour configurations for the best-performing models
of all different types. Note that the mixture model is aligned with the training
data by construction. All other models largely succeed in learning the
marginalized discrete likelihoods, with differences occurring mostly in
categories with low frequency.

\begin{figure}
    \includegraphics[width=1.\textwidth]{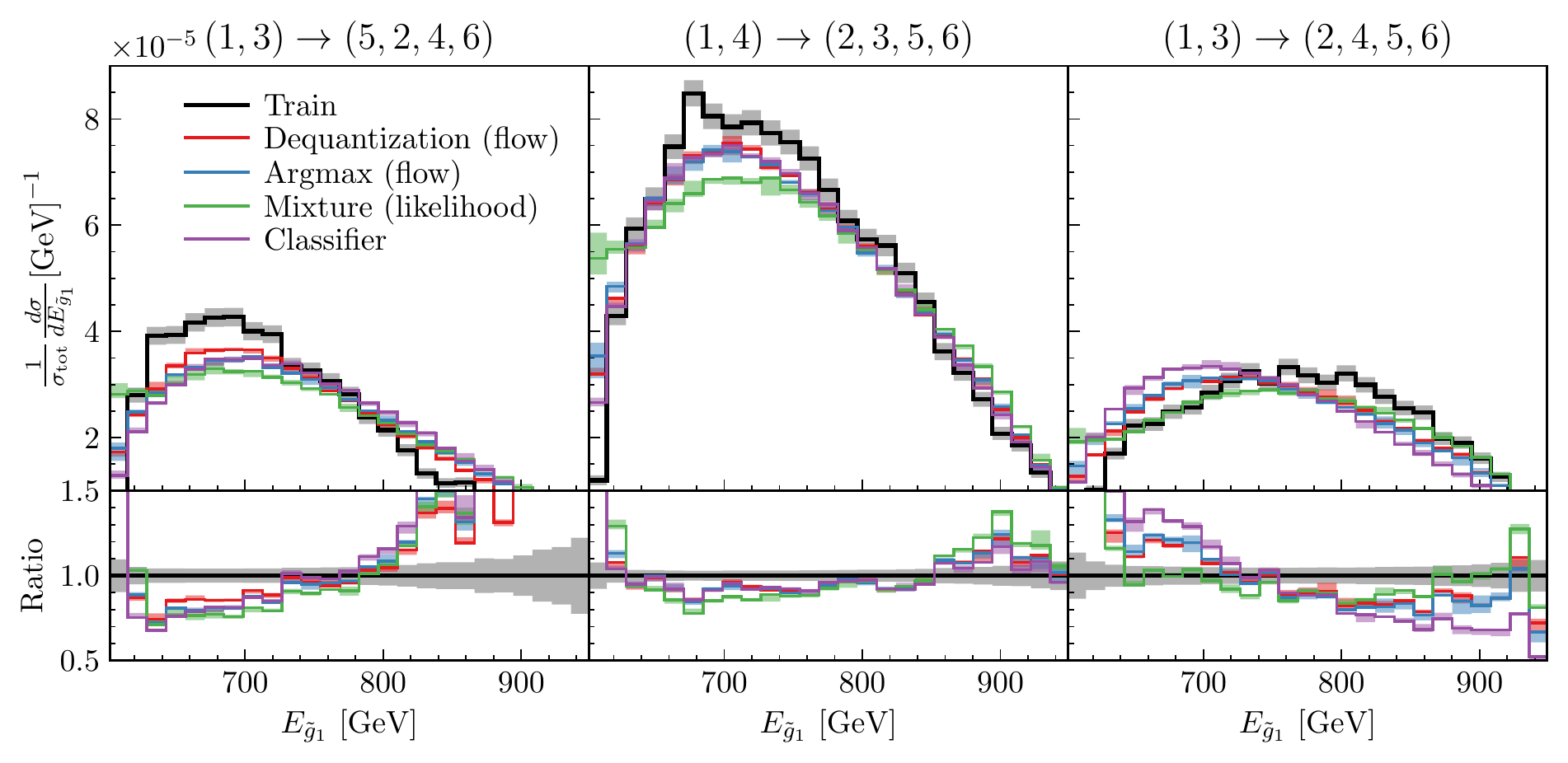}
    \caption{The gluino energy spectra of a selection of colour orderings as
    predicted by the same models as in figure \ref{fig:marginal-discrete}.}
    \label{fig:colour-discrete}
\end{figure}

\begin{figure}
    \includegraphics[width=1.\textwidth]{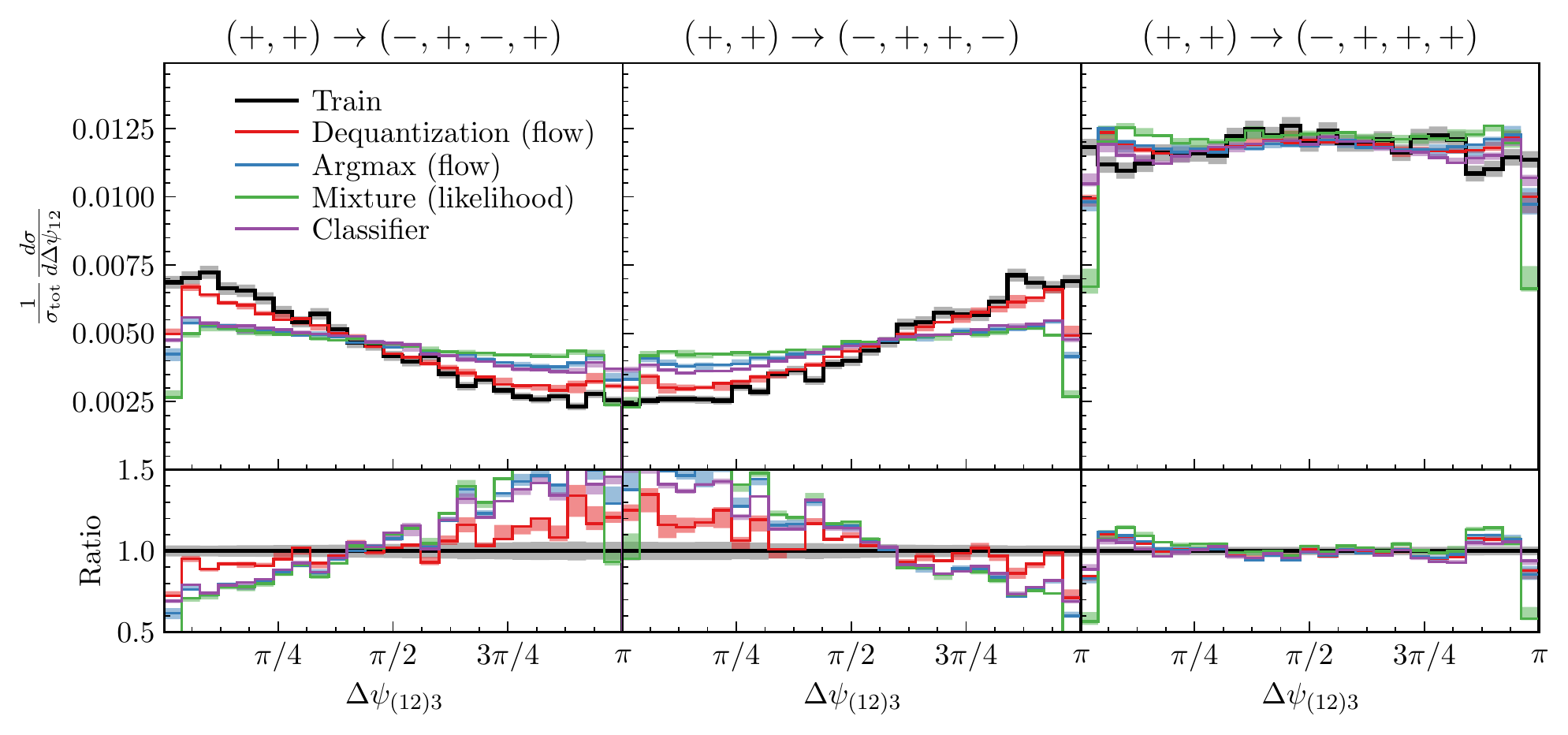}
    \caption{ The distribution of the $\Delta \psi_{(12)3}$ observable defined
        in the text of a selection of helicity configurations as predicted by
        the same models as in figure \ref{fig:marginal-discrete}.}
    \label{fig:helicity-discrete}
\end{figure}

Figures \ref{fig:colour-discrete} shows the gluino energy spectra for a
selection of colour orderings, while figure \ref{fig:helicity-discrete} shows
the distribution of the spin-sensitive observable $\Delta \psi_{(12)3}$ for a
selection of helicity configurations. The observable $\Delta \psi_{(12)3}$ is
defined as the angular separation between the plane spanned by $p_{\tilde{g}_1}$
and $p_{\tilde{g}_2}$, and the plane spanned by $p_{\tilde{g}_1} +
p_{\tilde{g}_2}$ and $p_{\tilde{g}_3}$. None of the models show substantially
better or worse performance. Note that the per-category training statistics are
quite limited. The performance of all models could likely be improved by the
techniques explored in
\cite{Diefenbacher:2020rna,Winterhalder:2021ave,Butter:2021csz}.

\section{Anomaly Detection} \label{sec:DM} One important application of ML-based
density estimation methods is their use as model-agnostic anomaly detectors.
Their objective is the identification of events that can be considered outside
the Standard Model (SM) density, and may thus point to new physics beyond
the SM (BSM). The use of an explicit SM density estimator has been successful in
a variety of anomaly detection tasks
\cite{Ostdiek:2021bem,Hallin:2021wme,Nachman:2020lpy,Buss:2022lxw,Jawahar:2021vyu,Dillon:2021nxw,Stein:2020rou,Caron:2021wmq}.
Here, we apply the methods set out in the previous section to the datasets of
the Dark Machines Anomaly Score Challenge \cite{Aarrestad:2021oeb}. In it,
numerous ML-driven methods were considered to identify anomalous events from a
variety of BSM models after being trained on SM background events.

Anomaly detection with an explicit likelihood estimator is accomplished by
relying on the principle that a model trained on SM events should assign small
likelihoods to out-of-distribution events. One thus defines an anomaly score as 
\begin{equation}
    s(x) = \frac{\log p(x) - \log p_{\text{min}}}{\log p_{\text{max}} - \log p_{\text{min}}},
\end{equation}
where $\log p_{\text{max}}$ and $\log p_{\text{min}}$ are respectively the
largest and smallest likelihoods assigned to the event samples evaluated in the
inference dataset. A cutoff point $s_{\text{cut}}$ can then be determined such
that events with $s(x_i) < s_{\text{cut}}$ are classified as anomalous. To
assess performance, one can then compute a background efficiency $\epsilon_B$
and a signal efficiency $\epsilon_S$ for a given value of $s_{\text{cut}}$,
which represent the fraction of events maintained after application of the cut.
Equivalently, the signal efficiency may be evaluated as a function of the
background efficiency, $\epsilon_S(\epsilon_B)$.

\subsection{Dataset} \label{sec:DM-data} The Dark Machines Anomaly Score
Challenge training datasets are derived from over 1 billion $13$ TeV simulated SM
LHC collisions. Through the application of different sets of cuts, four separate
channels were identified:
\begin{itemize}
    \item \textbf{Channel 1}: Hadronic activity with a lot of missing energy ($214$k events).
    \item \textbf{Channel 2a}: At least three identified leptons ($20$k events).
    \item \textbf{Channel 2b}: At least two identified leptons ($340$k events).
    \item \textbf{Channel 3}: Inclusive with moderate missing energy ($8.5$M events).
\end{itemize}
Furthermore, for testing purposes a variety of BSM signal events were generated
from a variety of models, such as those containing a $Z'$ and a collection of
supersymmetric models. Finally, a secret dataset is available with BSM signals
unknown to the challenge participants.

The data consists of events that are each composed of a missing transverse
energy $E_{T}^{\text{miss}}$ and its azimuthal direction
$\varphi_{T}^{\text{miss}}$, as well as a varying number of reconstructed
objects that are specified by their energy $E$, transverse momentum $p_T$,
pseudorapidity $\eta$, azimuthal angle $\varphi$, and object type: jet,
$b$-tagged jet, $e^+$, $e^-$, $\mu^+$, $\mu^-$ and $\gamma$.

\subsection{Models} \label{sec:DM-models}
The data described in section \ref{sec:DM-data} display all the
properties that were incorporated in the baseline flow model in section
\ref{sec:gluino}. The reconstructed objects adhere to permutation invariance and
can thus be either handled with a stochastic permutation or with a sort
surjection as described in section \ref{sec:perm-invariance}. Furthermore, the
number of objects is variable, which can be dealt with using the techniques of
section \ref{sec:varying-dim}. Finally, the object type represents a categorical
feature which can be dealt with using any of the methods of section
\ref{sec:discrete}. 

We preprocess the training data by normalizing all the features to zero mean and
unit standard deviation. Since the phase space is not as straightforwardly
constrained as in the case of the matrix element-level case of section
\ref{sec:gluino}, we opt to replace the multivariate uniform base distribution
of the baseline normalizing flow by a multivariate standard Gaussian. The number
of objects in the feature space is restricted to a fixed $N_{\text{max}}$,
meaning that the $N_{\text{max}}$ objects with the largest $p_T$ are included.
Including $E_{T}^{\text{miss}}$ and $\varphi_{T}^{\text{miss}}$, the
dimensionality of the baseline normalizing flow is $2 + 4 N_{\text{max}}$. The
events with fewer than the maximum number of objects have a number of empty
slots, which are filled with \texttt{NaN}s. As described in section
\ref{sec:varying-dim}, the baseline normalizing flow passes these \texttt{NaN}s
without affecting them\footnote{During the evaluation of the MADE network and
the classifier described in section \ref{sec:factorized}, \texttt{NaN} features
are set to zero.}, and a dropout layer is included immediately after the base
distribution to handle them. The hyperparameters of the baseline normalizing
flow remain the same (table \ref{tab:gluino-hyper}). To handle the categorical
features, we consider the following three models:

\begin{itemize}
    \item \textbf{Dequantization}: The categorical features are included using
    uniform dequantization as described in section \ref{sec:deq}. The object
    types are mapped to integer numbers in $[0,6]$, which are dequantized into
    the range $[-3.5, 3.5]$. This brings the total flow dimensionality to $2 + 5
    N_{\text{max}}$.
    \item \textbf{Mixture}: The categorical features are combined with the
    categorical distribution used for the dropout as described in section
    \ref{sec:factorized}. That is, the object type is mapped to an integer in
    $[0,7]$, where now $0$ means the absence of an object and $[1,7]$ are the
    existing object types. 
    \item \textbf{Classifier}: The discrete features are included through a
    separate classifier as described in section \ref{sec:factorized} with
    identical architecture as the one used in section
    \ref{sec:discrete-experiments}. 
\end{itemize}
All the above models include either a stochastic permutation or a sorting
surjection. We set $N_{\text{max}} = \{8,10\}$ for the dequantization and
classifier models, but restrict the mixture model to $N_{\text{max}} = \{6,8\}$
due to the quick proliferation of categorical configurations as the number of
objects increases.

\subsection{Results}
We evaluate the performance of the models described in section
\ref{sec:DM-models} on the four channels described in section \ref{sec:DM-data}.
An often-used measure of performance is the area under the curve (AUC) of the
receiver operating characteristic (ROC) curve, which shows the relationships
between the background efficiency $\epsilon_B$ and the signal efficiency
$\epsilon_{S}$. However, the AUC is dominated by the model performance at large
background efficiency, while in the anomaly detection context the performance at
small background efficiencies is often more relevant. Viewing the signal
efficiency as a function of the background efficiency, $\epsilon_S(\epsilon_B)$,
the performance metric proposed in \cite{Aarrestad:2021oeb} is the maximum
signal improvement
\begin{equation} \label{eq:max-SI}
    \text{Max SI} = \max_{\epsilon_B} \epsilon_S(\epsilon_B) / \sqrt{\epsilon_B}, \text{ where } \epsilon_B \in \{10^{-2}, 10^{-3}, 10^{-4} \}. 
\end{equation}

\begin{figure}
    \includegraphics[width=1.0\textwidth]{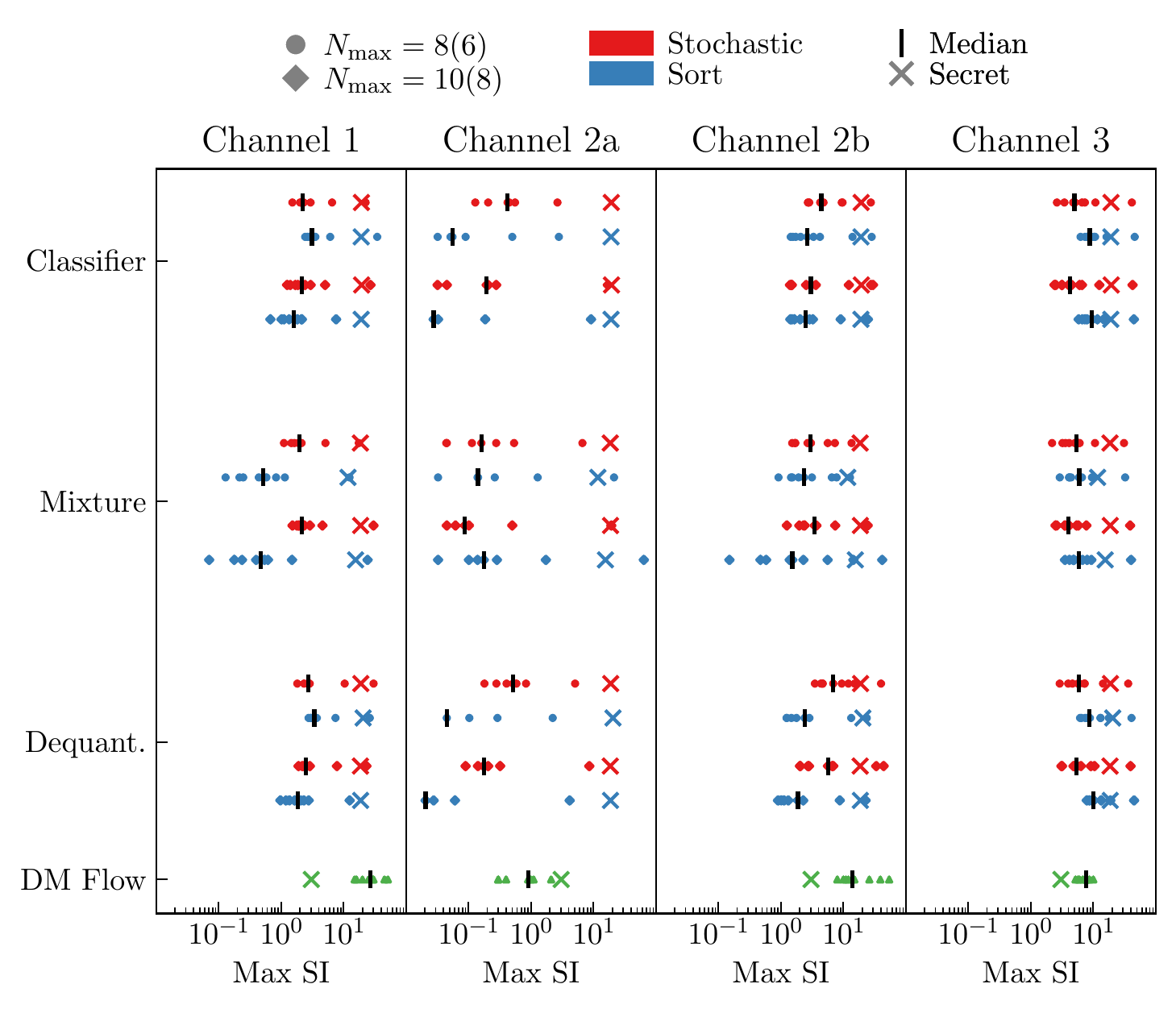}
    \caption{Max SI scores (eq.~\eqref{eq:max-SI}) for the three models listed
    in section \ref{sec:DM-models}, evaluated with stochastic permutation (red)
    and sort surjection (blue) and varying $N_\text{max}$ indicated by the
    marker shapes. The results of the flow model used in \cite{Caron:2021wmq}
    are shown in green diamonds. The black bar indicates the median performance
    of the BSM signals. The performance on the secret dataset is indicated with a cross.} 
    \label{fig:DM-results}
\end{figure}

Figure \ref{fig:DM-results} shows the per-channel max SI scores of the models
described in section \ref{sec:DM-models} and of the flow model used in
\cite{Caron:2021wmq}, evaluated on the test signals and on the secret dataset.

We first focus on the test signals. As expected, in almost all cases the models
with stochastic permutation outperform those with sort surjection in channels 1,
2a and 2b which are all limited by training statistics. On the other hand,
channel 3 provides much more data, which benefits the sort surjection models.
Similarly, models with larger $N_\text{max}$ tend to perform worse than their
counterpart with smaller $N_\text{max}$ in channels 1, 2a and 2b, again due to
limited training data. In channel 3 the performance is very similar, indicating
that the softest objects in the event are less relevant than the hard ones for
the purposes of distinguishing BSM signals from SM background. Among the
treatments of the discrete features, we find that the mixture model
underperforms in all channels. This is likely caused by the proliferation of
possible discrete configurations which all receive their own embedding, and
would thus likely require much more training data. Note that the stochastic
permutation transform does not offer much help in this specific case, because it
causes many new categories to appear that were not present in the sorted data.

When comparing to the flow model of \cite{Caron:2021wmq}, which was one of the
best-performing models \cite{Aarrestad:2021oeb}, we find a varied picture. The
model of \cite{Caron:2021wmq} significantly outperforms the models considered
here in the low statistics channels, while sorted classifier and dequantization
models outperform it in channel 3. There are several differences between the
models, but the most substantial one is the treatment of differing numbers of
objects. While the models considered here use the method considered in
\ref{sec:varying-dim}, the model in \cite{Caron:2021wmq} used a
dequantization-like procedure, where the features of missing objects are filled
by out-of-distribution noise. This procedure leads to non-vanishing bound
looseness, but it appears to provide better average performance for low
statistics. On the other hand, in cases where large amounts of training data are
available the method of \ref{sec:varying-dim} appears to be preferred due to its
access to the exact likelihood. 

Shifting our attention to the secret dataset, the results differ significantly.
The model of \cite{Caron:2021wmq} underperforms on the secret dataset in
channels 1, 2a and 3, while for essentially all models considered here the
performance on the secret data is consistently high. In fact, in channel 3 all
of them outperform the best-scoring models considered in
\cite{Aarrestad:2021oeb}, and some of them obtain the best performance in
channel 2a. The secret dataset consists of a wide variety of signals including,
for instance, fully unphysical events. It appears that the current models are
better suited to detect such anomalies.

\section{Conclusions} \label{sec:conclusions}
Normalizing flows have shown great promise in their application in various areas
of particle physics due to their simultaneous capabilities as event generators
and density estimators. However, their architecture does not provide much
flexibility in terms of modelling peripheral features that are commonly
associated with collision events. In this paper we explored, among other things,
the addition of surjective and stochastic transforms as part of the usual
normalizing flow architecture with the goal of increasing its flexibility.

In section \ref{sec:gluino} we considered the matrix element-level process $g
\kern 0.05em g \rightarrow \tilde{g} \kern 0.05em \tilde{g} \kern 0.05em
\tilde{g} \kern 0.05em \tilde{g}$ which displays four-fold permutation symmetry
and rich discrete colour and spin spectra. We explored enforcing permutation
symmetry through a stochastic permutation transform or a sort surjection, and
found that both are beneficial, but the correct choice depends on the available
training data. To incorporate the discrete features, we considered two
surjective transforms in the form of variational dequantization and an argmax
surjection, as well as two alternatives in the form of factorized models. We
find that the exact likelihood evaluation offered by the factorized model leads
to better performance. Finally, we also considered the issue of varying
dimensionality and introduce a surjective transform with vanishing bound
looseness to handle this situation.

In section \ref{sec:DM} we applied these techniques to the objective of anomaly
detection in the context of the Dark Machines Anomaly Score Challenge
\cite{Aarrestad:2021oeb}, comparing to the performance of the flow model used in
\cite{Caron:2021wmq} which was one of the best-performing models. We find
results that are largely consistent with the conclusions of section
\ref{sec:gluino}, and achieve substantially better results on the secret
dataset, outperforming all models considered in \cite{Aarrestad:2021oeb} in
channels 2a and 3.

We believe that the techniques and the assessment of their application to
typical collision events presented here will help improve future generative
modelling and density estimation. While many practical applications of
normalizing flows have already been explored, we expect that their general
applicability will find many more use-cases in the future.

\section*{Acknowledgements}
I would like to thank Melissa van Beekveld for evaluating performance on the
Dark Machines Anomaly Score Challenge secret dataset. This work was supported by
the European Research Council (ERC) under the European Union’s Horizon 2020
research and innovation programme (grant agreement No. 788223, PanScales).

\bibliography{refs}

\nolinenumbers

\end{document}